\documentclass[usenatbib]{mnras} 
\usepackage{txfonts}
\usepackage{bm}

\usepackage{graphicx,times}	

\title[The dynamical state of A2399]{The dynamical state of Abell 2399: a bullet-like cluster}
\author[Ana C. C. Louren\c{c}o et al.]{%
Ana C. C. Louren\c{c}o,$^{1,2}$\thanks{E-mail: ana.lourenco@postgrado.uv.cl }
P. A. A. Lopes,$^{2}$
T. F. Lagan\'a,$^{3}$
R. S. Nascimento,$^{4}$ 
\newauthor
R. E. G. Machado,$^{5}$
M. T. Moura,$^{5}$
Y. L. Jaff\'e,$^{1}$
A. L. Ribeiro,$^{6}$
B. Vulcani,$^{7}$
A. Moretti,$^{7}$
\newauthor
and L. A. Riguccini$^{2}$ 
\\
$^{1}$Instituto de F\'isica y Astronom\'ia, Universidad de Valpara\'iso, Valpara\'iso, Avda. Gran Breta\~na 1111 Valpara\'iso, Chile\\
$^{2}$Observat\'orio do Valongo, Universidade Federal do Rio de Janeiro, Ladeira do Pedro Ant\^onio 43, Rio de Janeiro, RJ, 20080-090, Brazil\\
$^{3}$N\'ucleo de Astrof\'isica, Universidade Cruzeiro do Sul/Universidade Cidade de S\~ao Paulo, Galv\~ao Bueno 868, Liberdade, 01506-000, S\~ao Paulo,SP, Brazil\\
$^{4}$Laborat\'orio Nacional de Astrof\'isica/MCTI, Itajub\'a, MG, 37504-364, Brazil\\
$^{5}$Departamento Academico de F\'isica, Universidade Tecnol\'ogica Federal do Paran\'a, Sete de Setembro 3165, Curitiba, PR, 80230-901, Brazil\\
$^{6}$Laborat\'orio de Astrof\'isica Te\'orica e Observacional, Universidade Estadual de Santa Cruz, Ilh\'eus, BA, 45650-000, Brazil\\
$^{7}$INAF-Osservatorio astronomico di Padova, Vicolo Osservatorio 5, IT-35122 Padova, Italy}
\date{Accepted 2020 August 12. Received 2020 August 12; in original form 2020 May 12.}

\pubyear{2020}

\begin{document}
\label{firstpage}
\pagerange{\pageref{firstpage}--\pageref{lastpage}}
\maketitle

\begin{abstract}
 While there are many ways to identify substructures in galaxy clusters using different wavelengths, each technique has its own caveat. In this paper, we conduct a detailed substructure search and dynamical state characterisation of Abell 2399, a galaxy cluster in the local Universe ($z \sim 0.0579$), by performing a multi-wavelength analysis and testing the results through hydro-dynamical simulations. In particular, we apply a Gaussian Mixture Model to the spectroscopic data from SDSS, WINGS, and Omega WINGS Surveys to identify substructures. We further use public \textit{XMM-Newton} data to investigate the intracluster medium (ICM) thermal properties, creating temperature, metallicity, entropy, and pressure maps. Finally, we run hydro-dynamical simulations to constrain the merger stage of this system. The ICM is very asymmetrical and has regions of temperature and pressure enhancement that evidence a recent merging process. The optical substructure analysis retrieves the two main X-ray concentrations. The temperature, entropy, and pressure are smaller in the secondary clump than in the main clump. On the other hand, its metallicity is considerably higher. This result can be explained by the scenario found by the hydro-dynamical simulations where the secondary clump passed very near to the centre of the main cluster possibly causing the galaxies of that region to release more metals through the increase of ram-pressure stripping.

\end{abstract}




\section{INTRODUCTION}
\label{sec:lit}

    The standard hierarchical model implies that galaxy clusters are formed by the accretion of galaxies and groups of galaxies \citep{PressSchechter1974}. Observational evidence at different wavelengths shows that this process continues to occur in the local universe \citep[e.g.][]{Hurier2017,Hurier2019} and galaxy clusters are still accreting groups and even suffering major mergers. Cosmological simulations show that a massive ($\sim$10$^{15} h^{-1} {\rm M}_{\odot}$) cluster at $z = 0$ has typically accreted $\sim$40\,per cent of its galaxies ($>10^9 h^{-1} {\rm M}_{\odot}$ ) from groups (more massive than $10^{13} h^{-1} {\rm M}_{\odot}$) \citep{McGee_2009}. Observations further suggest that 10--20\,per cent of low-$z$ clusters are undergoing cluster-cluster mergers \citep{Katayama_2003,Sanderson_2009,Hudson_2010}, and therefore, have not yet reached the relaxed dynamical state. 
    
    The identification of unrelaxed clusters is essential to investigate the possible influence they exert on the scattering of scaling relations \citep{Ventimiglia2008} and the uncertainties of cosmological parameters obtained through galaxy cluster studies \citep{Dooley2014}. Also, that is fundamental to understand how interactions impact the evolution of galaxies \citep{Stroe2014, Kelkar_2020}. However, conventional observational methods to study cluster dynamics are limited. 
    
    An important issue in the assessment of a cluster dynamical state from optical data is that we can normally only measure projected quantities. Ideally, one would use 3D information, instead of only RA and Dec. Redshift surveys help mitigate this problem, but those should have high completeness so that the identification of substructures and interactions is reliable. Even so, radial velocity has been used as a proxy for a third spatial coordinate in several substructure studies \citep[e.g.][]{Einasto2010,Ribeiro2013,Cohen2014,Jaffe2016}. 
    
    In the case of the X-ray substructure studies, important information about the dynamical state of the gas could be derived from the analysis of spectral maps, instead of using solely the surface brightness map. Recently, in a sample of 53 galaxy clusters, \citet{Lagana_2019} found that 16 clusters classified as cool-core \citep[CC, a drop in the temperature profile towards the centre that is common in relaxed clusters;][]{2015_Ichinohe} by \citet{Andrade_Santos_2017} according to concentration, cuspiness, and central density of the surface brightness map have instead characteristics that indicate interactions such as asymmetries and discontinuities in their temperature maps. 
    
    Commonly, cases of cluster mergers that are not obvious in the optical data are clearly seen in the analysis of the X-ray images. The interactions between clusters are extremely powerful phenomena that release a high amount of energy \citep[of the order of $10^{64}$\,erg, see ][]{Sarazin2002}. Such events can cause the Intracluster Medium (ICM) to exhibit large deviations from spherical symmetry \citep{Roettiger_1996}, an increase of the temperature in some specific regions via gas compression \citep{Markevitch_2002,Markevitch_2005,Russell_2010,Macario_2011,Owers_2011,Russel_2012}, destruction of the cool-core, and a displacement between the Brightest Cluster Galaxy (BCG) and X-ray peak \citep{Lopes_2018}. 
    
    However, since the thermal bremsstrahlung emission of the ICM depends quadratically on the gas density, not all galaxy clusters are luminous enough in X-rays to allow a substructure study at large clustercentric radii; to worsen the situation, the gas can also suffer from projection effects that make it virtually impossible to identify substructures in the line of sight since the ICM is optically thin.
    
    Numerical simulations have often been employed to get a deeper view of cluster mergers \citep[e.g.][]{ZuHone2010, ZuHone2011, ZuHone2013a, Vazza2012, Iapichino2017, Schmidt2017}. In particular, they are also applied to model dissociative merging clusters, in which the gas has been detached from the dark matter peaks, helping to constrain parameters such as the mass ratio, angle of inclination of the collision with respect to the line of sight, and also, the velocity of propagation of the shock fronts in the ICM \citep[e.g.][]{Springel2007, Mastropietro2008, vanWeeren2011, Machado2013, Lage2014, MachadoMonteiro2015, Monteiro2017}. Furthermore, numerical simulations predict that interacting clusters experience an increase in the ram pressure that can originate gas stripped galaxies with a jellyfish morphology \citep[e.g.][]{Vijayaraghavan2013, Ruggiero2019}. This enhancement occurs because the ram pressure grows with the square of the relative velocity between the intracluster medium and galaxies which is higher in merging clusters. Merger events also increase the density of the ICM driving an increase in the ram pressure as well.  
    
    In this paper, we study in detail the dynamical state of Abell 2399, a galaxy cluster located at a filamentary intersection of the Aquarius-Cetus supercluster at $z = 0.0579$ \citep{Bregman2004}. It was classified as BM Type III by \citet{Bautz1970}, meaning that there are no central dominating (cD) cluster galaxies. $M_{200}$ and $L_{X}$ are respectively $4.31\,\times\,10^{14}\,{\rm M}_{\odot}$ and $0.636\,\times\,10^{44}$\,erg\,s$^{-1}$ \citep{Lopes_2009_b}. This galaxy cluster has been the subject of several studies on its dynamical state obtaining conflicting conclusions summarised bellow.
    
    Abell 2399 was selected in this paper for presenting traces of multi-modality in its velocity distribution and in the plane of the sky. In addition, its temperature profile is not isothermal, suggesting that some non-gravitational processes, such as an interaction, could be playing a role here. The X-ray morphology of A2399 is suggestive of a possible collision event \citep[e.g.][]{Bohringer2010, Mitsuishi2018} and from the inspection of its optical data, we found more than one possible BCG, which is another indication of interaction (see Fig.~\ref{fig:multiwavelength}).
    
    Some authors have classified A2399 as a single cluster, with no interactions. \citet{Flin2006}, for example, applied a transform wavelet code in an attempt to detect substructures in the projected distribution of the galaxies of A2399 and other 182 Abell clusters. In order to do that they first combined data from the 10-inch photographic plates taken with the 48-inch Palomar Schmidt Telescope and from the Digitized Sky Survey. Their code did not detect any substructure in this 2D analysis and the system was classified as uni-modal. 
	
	Abell 2399 was also studied as part of the WIde-field Nearby Galaxy clusters Survey (WINGS) which had as the main goal to investigate the galaxy populations in clusters in the local universe ($0.05 \le z \le 0.07$) and the environmental effects on their stellar populations. This survey obtained spectroscopic data for 48 clusters including A2399. On the investigation of substructures in the WINGS clusters, \citet{Ramella2007} applied a kernel-adaptive code, also in the spatial projected distribution of galaxies. In agreement with \citet{Flin2006}, they found no substructures in the A2399 two-dimensional analyses.
	
	The same conclusion was obtained by \citet{Moretti2017}, who applied a $\pm 3\sigma$ clipping method to assign membership to the OmegaWINGS sample. Their analysis did not indicate another structure separated in the line of sight of this system.

	On the other hand, other authors have found evidence in different wavelengths that A2399 is a cluster merger. \citet{Bohringer2010} inspected the X-ray emission of A2399 (or RXCJ2157.4-0747) with good details in the \textit{XMM-Newton} image. The authors identified two well-separated components within $R_{500}$ (11.1\,\mbox{arcmin}). Contrary to the classifications derived from the optical data available at that time, they classified this cluster as bi-modal. It is noteworthy that A2399 has very diffuse, low surface brightness regions in its centre that make it difficult to automatically identify its local maximum \citep{Bohringer2010}. 

	In the same direction, \citet{Fogarty2014} reported a decrease in the fraction of elliptical galaxies with a slow rotation towards the denser regions of the cluster, while the two central galaxies are classified as having a fast rotation (one of them is the brightest cluster galaxy, BCG). They then suggested that the slow-rotation galaxies belong to infalling groups and are entering the cluster for the first time. They also state that this is the most complicated cluster in their sample of three clusters and that it is most likely to be a merger.

	More recently, \citet{Owers2017} used data from the Sydney-AAO Multi-Object integral field Spectrograph Galaxy Survey (SAMI-GS) to study substructures in galaxy clusters. When performing tests of substructures in 1D (velocity), 2D (sky coordinates), and 3D (velocity and sky coordinates), they found evidence of substructure in all three dimensions. They also found more than one candidate for the central cluster galaxy (CCG) and that the galaxy closest to the centre of the galaxy density distribution is not the BCG.
	
	Analysing the \textit{XMM-Newton} and \textit{Suzaku} images together, \citet{Mitsuishi2018} found several pieces of evidence for a merger scenario in the plane of the sky. Among them are the offset between temperature peaks and X-ray emission peaks, a significant difference between the location of the optical substructures and X-ray clumps, and discontinuity in X-ray surface brightness and temperature profile in the secondary clump with the possibility of a cold front. According to them, a merger scenario also explains the high entropy in the central region of the cluster found by the authors. In summary, substructure analyses on A2399 have been contradictory, making of this cluster an interesting object for further analysis and determining its dynamical state.
    
    The main goal of this paper was to probe the dynamical state of A2399 circumventing the weaknesses of the optical and X-ray merger identification techniques alone. In order to do that, we applied a careful substructure analysis combining optical and X-ray data to this merger candidate cluster. Optical substructures were identified by using a Gaussian Mixture Model (GMM), an algorithm that assumes that all the data points are generated from a mixture of a finite number of Gaussian distributions \citep{dudaHart1973}, in the high-completeness spectroscopic data from SDSS-DR12 \citep{Alam2015}, WINGS \citep{Cava_2009}, and Omega WINGS Surveys \citep{Moretti2017}. Then, we compared the results with the distribution of the thermodynamical properties of the gas. The parameters obtained from both analyses were used as input for hydrodynamical numerical simulations that we performed with the purpose of reproducing the observed temperature map and gas distribution. We intended to test the hypothesis that those properties of A2399 might be explained by a bullet-like collision.
    In Section~\ref{sec:sample} we present details about the sample and data reduction in the optical and X-ray. In Section~\ref{sec:substructure_ident} we perform a careful substructure analysis in the optical and X-ray and then compare both analyses. Finally, in Section~\ref{sec:hid_sim} we use the physical parameters obtained from the optical analysis to test if a bullet-like collision scenario would reproduce well the temperature map of the ICM obtained by us in this work.
    
    Throughout this paper, we adopt a $\Lambda$CDM cosmology with $H_{0} = 70\,{\rm km}\,{\rm s}^{-1}\,{\rm Mpc}^{-1}$, $\Omega_{\rm M} = 0.27$, $\Omega_{\Lambda} = 0.73$.
    Error bars are quoted within 1$\sigma$.
    
\section{SAMPLE AND DATA REDUCTION}
\label{sec:sample}

    We analyse optical and X-ray data available in the region covering the galaxy cluster A2399 in order to look for substructures and clues to its past merger history. The cluster was selected as an interacting candidate cluster after visually inspecting the velocity distribution along with the projected position vs.~velocity phase space (PS) diagram of the cluster galaxies retrieved from the supplemental version of the Northern Sky Optical Cluster Survey \citep[NoSOCS,][]{Lopes2004,Lopes2009}. The projected PS diagram is a schematic way of analysing galaxies in the different cluster's regions (infalling, virialized, and backsplash) and understanding the assembly history of galaxy clusters \citep{Jaffe_2015,Jaffe2016,Rhee2017}. The distribution of the cluster galaxies in single virialized clusters has a shape that resembles a trumpet \citep{Diaferio_1997,Diaferio_1999,Oman_2013,Haines_2015}. When searching for interacting candidates for further analysis, we looked for PS diagrams that have groups of galaxies outside of this trumpet region. 
  
\subsection{Optical Data}
\label{subsec:opt_sample}

    The spectroscopic follow-up program WIde-field Nearby Galaxy-cluster Survey (WINGS-SPE) is a large survey that obtained 48-cluster multi-fibre spectroscopy between $z = 0.04 - 0.07$ in an intermediate resolution of 6--9\,\AA \citep{Cava_2009}. Its parent sample, the WINGS-OPT, consists of 77 galaxy clusters, with optical imaging, in $B$ and $V$ bands, X-ray selected in a distance from the galactic plane of \citep[$b \geqslant 20^{\circ}$;][]{Fasano_2006}. Each cluster was observed in a field of view of $34' \times 34'$, $\sim$ 0.5 $R_{200}$ at $z = 0.05$. In order to extend the sample study to the infall regions of the clusters, the WINGS team observed 46 clusters of the original sample with OmegaCAM/VST imaging instrument in $u$, $B$, and $V$ bands covering $1^{\circ} \times 1^{\circ}$ of each cluster, $\sim$ 1.5 $R_{200}$ at $z = 0.05$ \citep{Gullieuszik_2015}. Recently the OmegaWINGS survey obtained spectroscopy of 33 clusters of its sample with the AAOmega spectrograph \citep{Smith_2004,Sharp_2006} with a spectral resolution of 3.5--6\,\AA full width at half maximum \citep{Moretti2017}.
    
	We retrieved spectroscopic data from WINGS and OmegaWINGS in a radius of 2.6\,Mpc, $\sim$1.68\, $R_{200}$ or $37.89'$ \citep[$R_{200}$ from][]{Biviano_2017}, around A2399's BCG, located at RA $= 329.\!\!^{\circ}37258$ and Dec. $= -7.\!\!^{\circ}79569$. We obtained 289 galaxies with 0.05 $\leq$ $z_{\rm spec}$ $\leq$ 0.07. 
		
	We also retrieved spectroscopic data from the Sloan Digital Sky Survey Data Release 12 \citep[SDSS-DR12;][]{Alam2015} within the same radius and redshift range we used to select the WINGS and OmegaWINGS data. Only the data with flags $\texttt{p.clean} = 1$ and $\texttt{s.z.warning} = 0$ were used. Those SDSS-DR12 flags ensure a dataset with reliable measurements. The first one removes duplicate objects, galaxies with deblending problems, interpolation problems, and suspicious detections such as saturated sources. The latter ensures that the redshift has been measured with a good S/N, has strong emission or absorption lines and that the largest part of the pixels is reliable so that we can rely on the redshift measure.

\begin{figure*}
      \includegraphics[width=8.5cm]{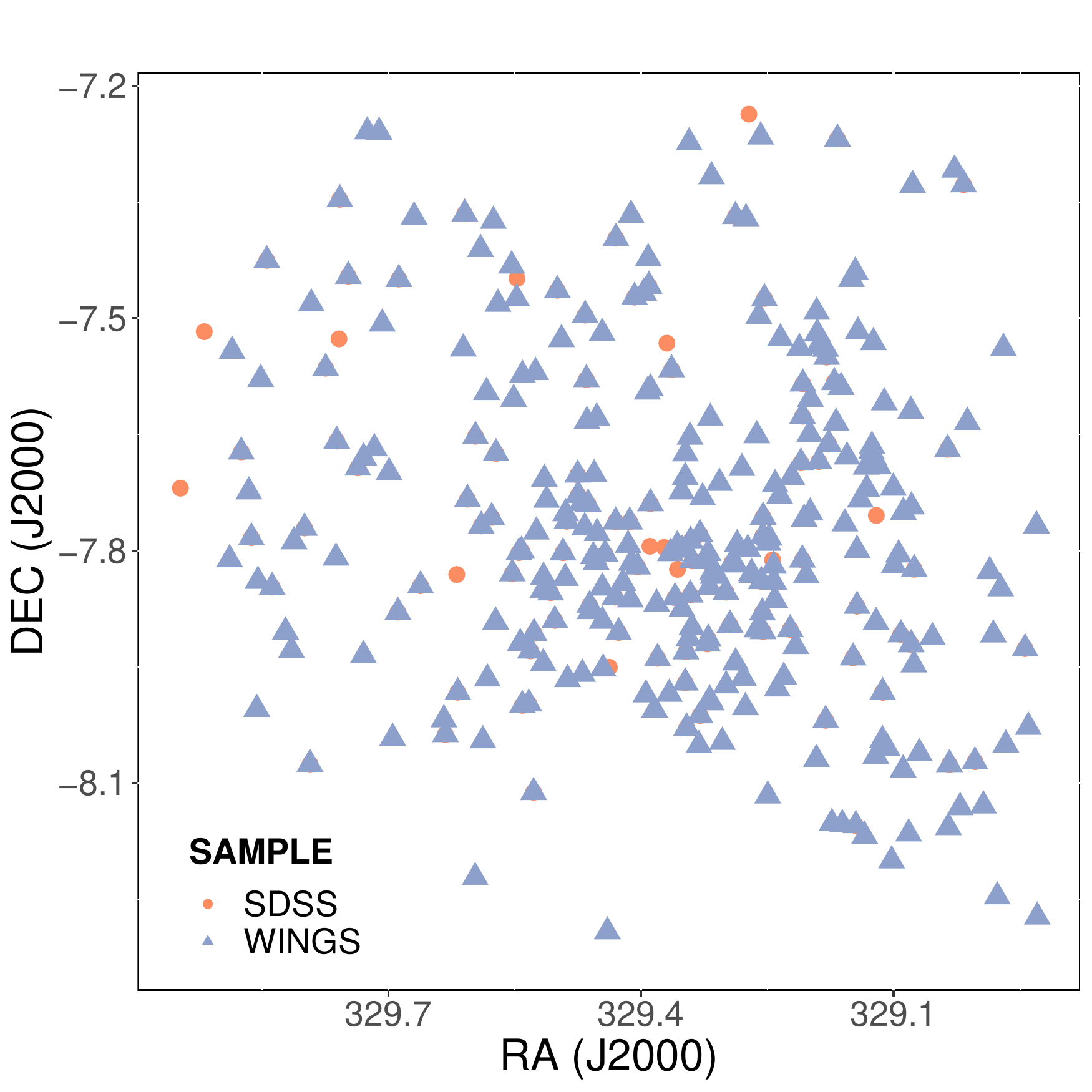}
      \includegraphics[width=8.5cm]{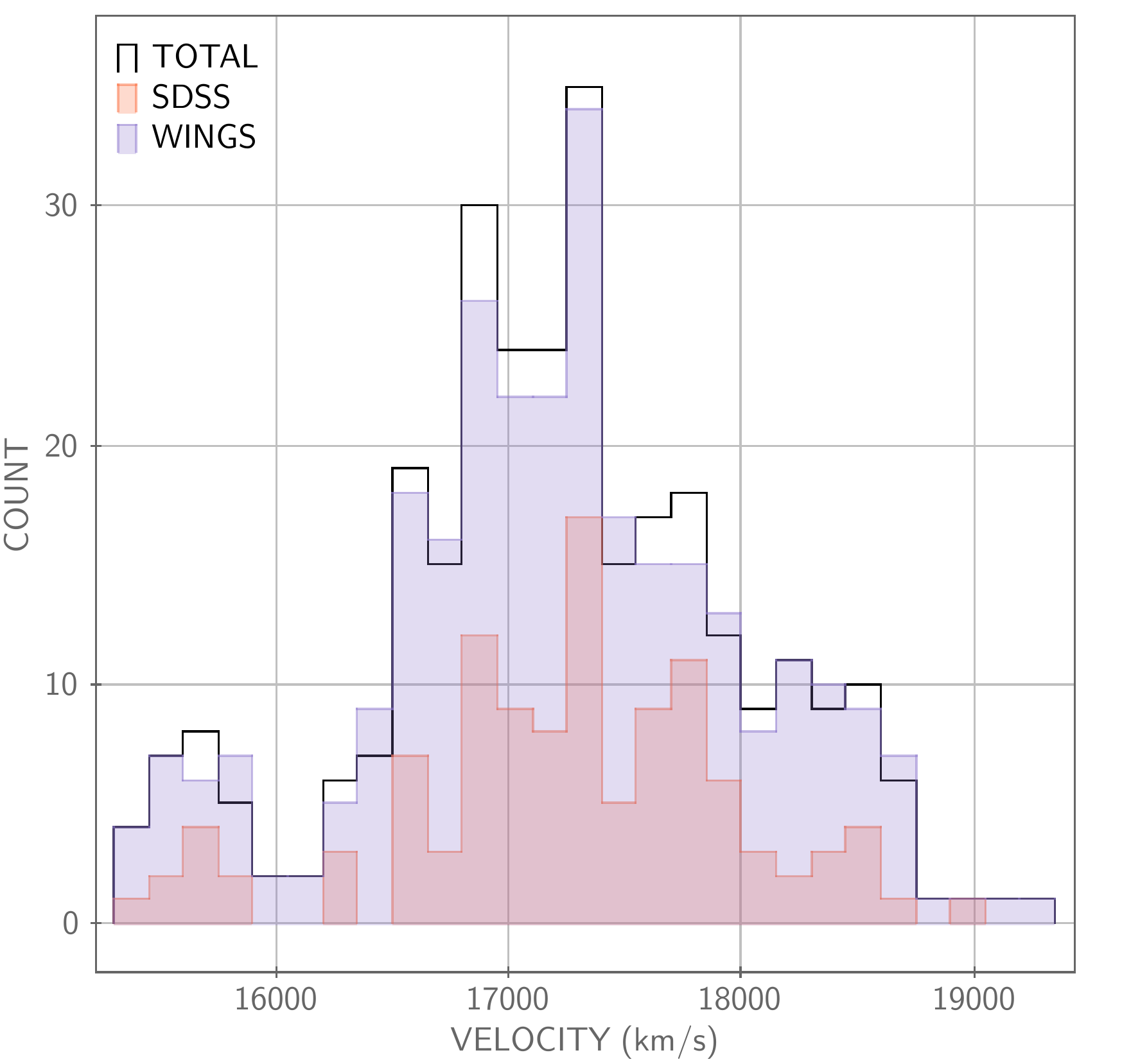}
      \caption{Left panel: The distribution of WINGS and SDSS samples in the plane of the sky. Right panel: The velocity distribution of the two matched samples, SDSS (red) and WINGS (blue), and the resulting sample (black). The duplicated galaxies from SDSS were excluded from our analysis. The SDSS sample alone has gaps in the velocity distribution that could bias the substructure identification.}
      
      \label{fig:sample}
\end{figure*}

    Although in SDSS-DR12 113 galaxies were retrieved, yet in the same region of WINGS and OmegaWINGS for equal redshift range, we obtained 289 galaxies with $z_{\rm spec}$. After performing a match between the two spectroscopic catalogues with a 1\,arcsec error, we obtained a combined sample of 299 galaxies. Duplicated galaxies from the SDSS catalogue were excluded from our analysis. Considering only WINGS/OmegaWINGS spectra, the spectroscopic completeness is 60.9 per cent (down to $m_{V}$ $=$ 20 and within a range of $B - V$ between 0.4 and 1.8). After including additional SDSS spectra (which reach a $m_{V}$ $\sim$18) the completeness increases to 69.3 per cent. The WINGS and OmegaWINGS spectroscopic data are 1.5 mag deeper than SDSS. This makes these surveys an appropriate source of data for the careful substructure analysis we performed in this work. The left panel of Fig.~\ref{fig:sample} shows the distribution of the WINGS and SDSS data in the plane of the sky, while in the right panel the velocity distribution of the two samples is shown along with the final combined sample without duplicate galaxies.  
    
\subsection{X-ray Data}
\label{subsec:xray_data}

    The \textit{XMM-Newton} observation (ID: 0654440101) was obtained on 2010 June 7. The thin 1 filter was used for all the cameras, in Prime full window for MOS and Prime full window extended for pn. 
    
    We reduced the data with the SAS version 17.0.0 (June 2018) and calibration files were updated to 2018, March. 
    
    In order to filter background flares, we applied a $2\sigma$ clipping procedure using the light curves in the [1--10]\,keV energy band. After cleaning the light curves, the resulting exposure times were approximately 53\,ks for MOS1, 59\,ks for MOS2, and 23\,ks for PN. Subsequently, the redistribution matrix file (RMF) and ancillary response file (ARF) were generated with the SAS tasks \textit{rmfgen} and \textit{arfgen} for each camera and for each region that we analysed. 
    
    Finally, we had to consider the background contribution of each detector. To account for this we obtained a background spectrum in an external annulus within 9 and 11\,arcmin away from the cluster centre, in the [10--12]\,keV energy band. Then, we compared these spectra with the one obtained from the \citet{Read2003} blank sky in the same region and energy band. The annulus background was divided by the blank sky background to obtain a normalisation parameter for each detector that was used in all spectral fits (as already presented in \citealt{Lagana2008}). 
    
    Point sources were detected by visual inspection, confirmed in the High Energy Catalogue 2XMMi Source, and excluded from our analysis.

\subsubsection{Spectral fit}

    The cluster’s X-ray emission was modelled in XSPEC v12.10.0 \citep{Arnaud1996}, with APEC single temperature plasma model \citep[][]{Smith2001} using the AtomDB v3.0.9 and WABS to model photoelectric absorption.
    
    We acquired the spectra in the [0.7--5]\,keV band, with the low-energy cut-off selected such as to diminish contamination from the residual soft emission of the Milky Way. The upper-temperature cut was made in order to increase the signal-to-noise ratio (S/N) since for low-temperature clusters, such as A2399 with $<kT>$ = 2.6\,keV \citep{Mitsuishi2018}, no emission is expected above this limit. The energy band from 1.3 to 1.9\,keV was excluded to prevent any influence from the strong Al and Si lines. 
    
    The redshift was fixed to $z = 0.0579$ \citep{Biviano_2017} for each fit, and the hydrogen column density, $nH$, was frozen to the value listed in the Leiden/Argentine/Bonn (LAB) value of 3.05 $\times$ $10^{20}\,{\rm cm}^{-2}$. We maintained the gas temperature ($kT$), metallicity ($Z$), and normalisation ($N_{\rm apec}$) as free parameters in the APEC plasma model. The metallicity was obtained from \citet{Asplund2009} in solar units.

\subsubsection{Projected maps}

	A detailed way of getting information on the cluster's history and dynamical state is obtaining spectral maps for different thermodynamic parameters. \citet{Lagana_2019} showed that very often clusters classified as cool-core have non-relaxed gas, and non-cool-core clusters have their gas relaxed. Therefore, the mere classification according to the cool-core presence is not sufficient for a reliable determination of the cluster dynamical state. It is then necessary to use detailed techniques such as spectral maps to understand the history of the cluster through comparison with hydrodynamic simulations, as we will see in the Section~\ref{sec:hid_sim}.
	
	To obtain the spatially resolved maps, we divided the data into small regions from which spectra were extracted. The 2D maps were made in a grid, where each pixel is 512 $\times$ 512 \textit{XMM-Newton} EPIC physical pixels, corresponding to 12.8\,arcsec $\times$ 12.8\,arcsec. We set a threshold of 900 counts per grid region after background subtraction to assure a signal-to-noise of at least 30. The spectra of the MOS1, MOS2, and pn detectors were simultaneously fitted and the best temperature and metallicity values were attributed to the central pixel of each grid \citep[as described in][]{Durret2010, Durret2011, Lagana2015, Lagana_2019}. It is important to highlight that since the spectral extraction regions are typically larger than the pixel map, individual pixel values are not independent. 

\section{SUBSTRUCTURES IDENTIFICATION}
\label{sec:substructure_ident}
\subsection{\textit{Mclust}}
\label{subsec:mclust}

	The \textit{mclust} method is a quite popular \textit{R} package which uses a Gaussian Mixture Model along with hierarchical clustering to determine the number of clusters in a dataset and to assign the data points memberships. In addition to classifying, it provides uncertainties and estimates the density-based finite Gaussian mixture modelling \citep{Scrucca_2016}.

	Gaussian mixture models are probabilistic models that fit a finite number of Gaussian distributions with unknown mean and covariance to the dataset. These models make use of an Expectation-Maximisation (EM) algorithm that acts similarly to $k$-means, but with the advantage of better-fitting asymmetric and overlapping datasets, since the only parameter used by $k$-means is the average and it uses Euclidean distances.
	
	At a basic level, the EM algorithm maximises the likelihood of the density estimation calculating at each iteration the expectation of the log-likelihood function using the current estimate of the parameters (Expectation step) and updating the parameters to maximise the log-likelihood function (Maximisation step). The first step of this algorithm considers fixed the parameters that describe the Gaussian and calculates for each point the probability of belonging to each cluster. In the second step, maximisation, the probability of each point belonging to a particular cluster is considered fixed and the Gaussian parameters are weighted by the sum of the probabilities of each point belonging to the clusters. The two steps of the EM increase the log-likelihood of the model which is the logarithm of the sum of the probability of each point belonging to a given Gaussian of the mixture. The iterations continue until converging.
	
	In summary, \textit{mclust} calculates the log-likelihood for each iteration and then computes the most effective approximation to estimate the mixtures by varying the shape, orientation, and volume of multidimensional Gaussians. \textit{Mclust} uses the Bayesian Information Criterion (BIC) to decide between the sixteen templates fitted by the code (see Appendix~\ref{appendix:a}). In the model fitting, the likelihood can be increased by adding parameters, but this may cause over-fitting and more groups than those that actually exist to be found. The BIC solves this issue by inserting a penalty term for the number of parameters in the model. Although, the model selection using the BIC cannot be done blindly because, for models with distinct numbers of free parameters that have their log-likelihood-ratio sequence bounded in probability, BIC follows the parsimony principle leading the probability of choosing the model with the fewer parameters to become higher as the number of parameters increases \citep{Findley91}.
	
	For the optical data, we performed a \textit{mclust} analysis on 1D on the velocity distribution, 2D on RA and Dec., and finally on 3D using RA, Dec., and velocity. Since RA and Dec. coordinates vary by a few degrees, while the velocity coordinates vary by thousands of kilometres per second, we applied the \textit{scale} function of \textit{R} to centre and scale the units in the 3D analysis. The modes identified by the \textit{mclust} in 3D were then isolated and had their centroids calculated for RA, Dec., and $z_{\rm spec}$. In this work, we chose to use the results of the \textit{mclust} 3D classification for the comparison with the ICM distribution. Unlike the 1D and 2D results, the groups found in the three-dimensional analysis are reliable in the sense that one is well distinguishable in the phase space diagram and the other seems to be associated with the western clump observed in the X-ray, as seen in Fig.~\ref{fig:opt_groups_xray}.
	
	One caveat of this method is that the data distribution is not necessarily described by a mixture of Gaussians. Not all the galaxies in the sample belong to the group that they were assigned by the GMM. In order to neutralise this forced membership attribution, we ran a \textit{shifting gapper} technique \citep{Fadda1996, Lopes2009} to identify objects that are not linked to any structure. In the end, we obtained the cluster's and the groups' members and galaxies that are cinematically in the region, but with low connection to the identified groups, the interlopers.
	
	Our \textit{mclust} analysis was performed in the range $cz = 14000-21000$\,km\,s$^{-1}$ within $\sim$1.5\,$R_{200}$ ($\sim$2.6\,Mpc) around the BCG in the combined SDSS-DR12, WINGS and OmegaWINGS datasets \citep[see][for examples of \textit{mclust} applied in substructure analyses]{Einasto_2012a,Einasto_2012b,Ribeiro2013,Monteiro_Oliveira_2020}.

\subsubsection{1D Analysis}
    In a relaxed cluster, a Gaussian velocity distribution can be assumed \citep[e.g.][]{Yahil1977}. The one-dimensional analysis of the velocity distribution has been used as the only source of information about the clusters' dynamical state in several studies \citep[see e.g.][]{Hou2009, Ribeiro2010, Ribeiro2011, Ribeiro2013}.
    
\begin{figure*}
      \includegraphics[width=9.0cm]{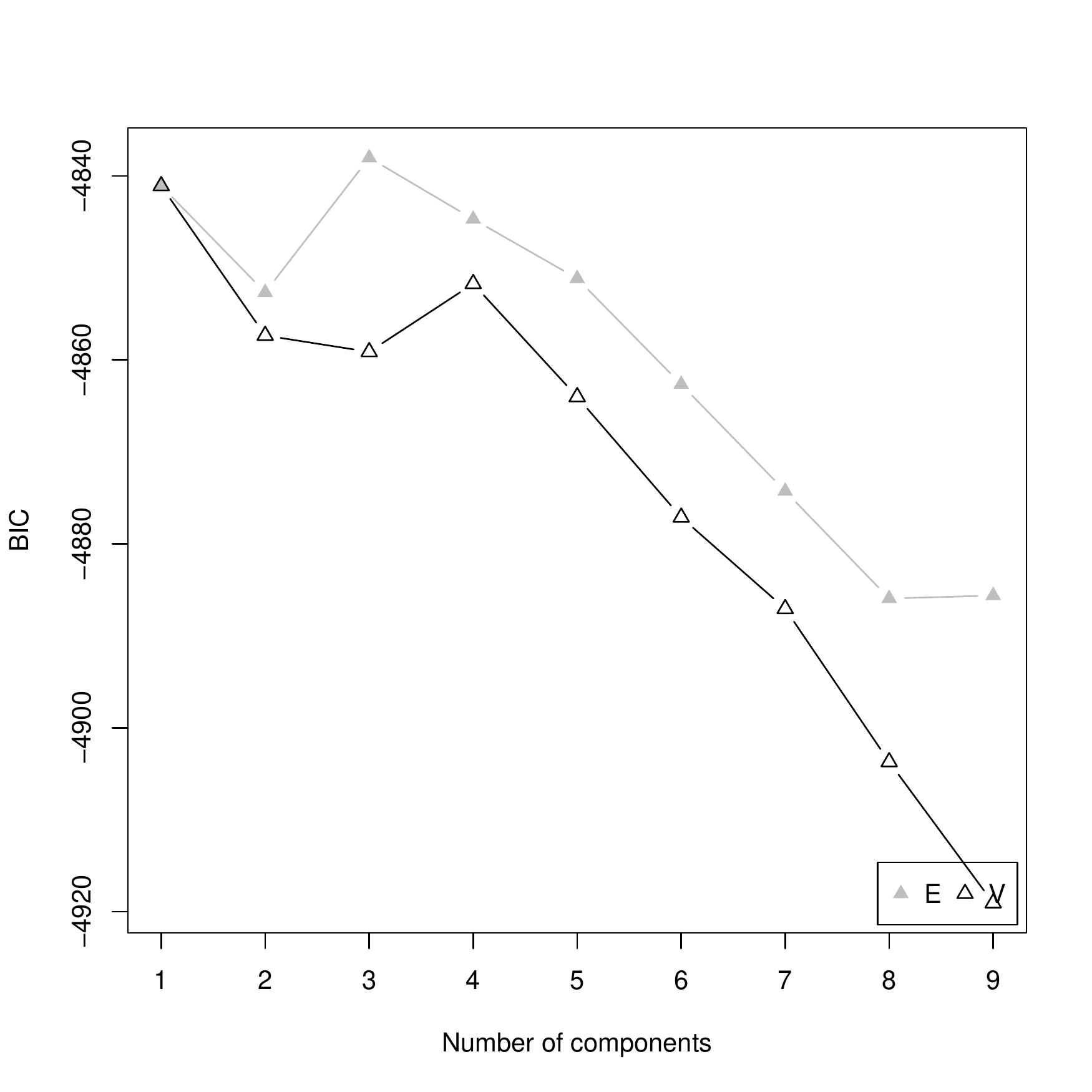}
      \includegraphics[width=8.5cm]{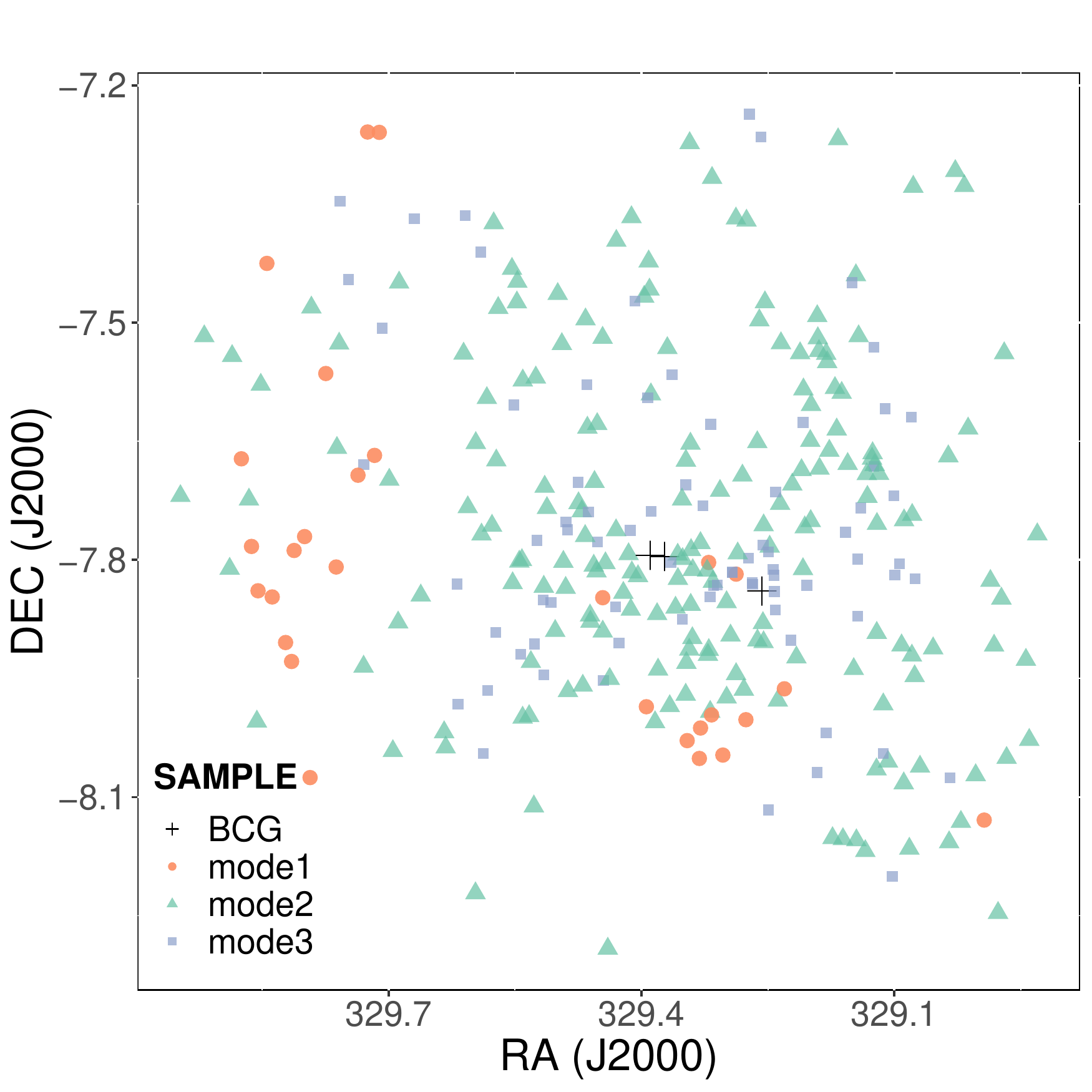}
      \caption{1D analysis. Left panel: BIC analysis for the different models fitted by \textit{mclust}. Model E with three velocity components is the one to better fit the data. Right panel: Distribution of the velocity groups found by the GMM in the plane of the sky. The black crosses mark the positions of the BCGs.}
      \label{fig:1D}
\end{figure*}

    Analysing only the velocity distribution, the code identified three velocity modes, suggesting a line of sight interaction scenario within a clustercentric radius of 2.6\,Mpc. The two best models fitted by \textit{mclust} in one dimension disagree. Model E, that fits Gaussians with equal variance, found three modes, while the variable variance model, V, found only one mode. The differences between the BIC of a uni-modal and a tri-modal distribution is smaller than 5. Hence, the number of clusters found by the chosen model is not reliable \citep{Kass1995}. The left panel of Fig.~\ref{fig:1D} shows the variation of the BIC according to the number of components fitted by the two different models. The model with the highest BIC is the one chosen. The right panel shows the distribution of the three velocity subgroups in the plane of the sky. The black crosses mark the position of the BCGs. It is noteworthy that the velocity groups found by the code present low cohesion when plotted in the plane of the sky. For example, mode 1 (red circles) has an isolated subgroup of galaxies closer to the centre of the cluster and another one more on the outskirts. This means that we cannot trust the substructure analysis using the velocity distribution only.
	
\subsubsection{2D Analysis}

    Another way to search for substructures in the optical is to perform an analysis of the distribution of galaxies in the plane of the sky looking for over-densities that may be associated with candidate clusters for interaction.
    
\begin{figure*}
      \includegraphics[width=9.0cm]{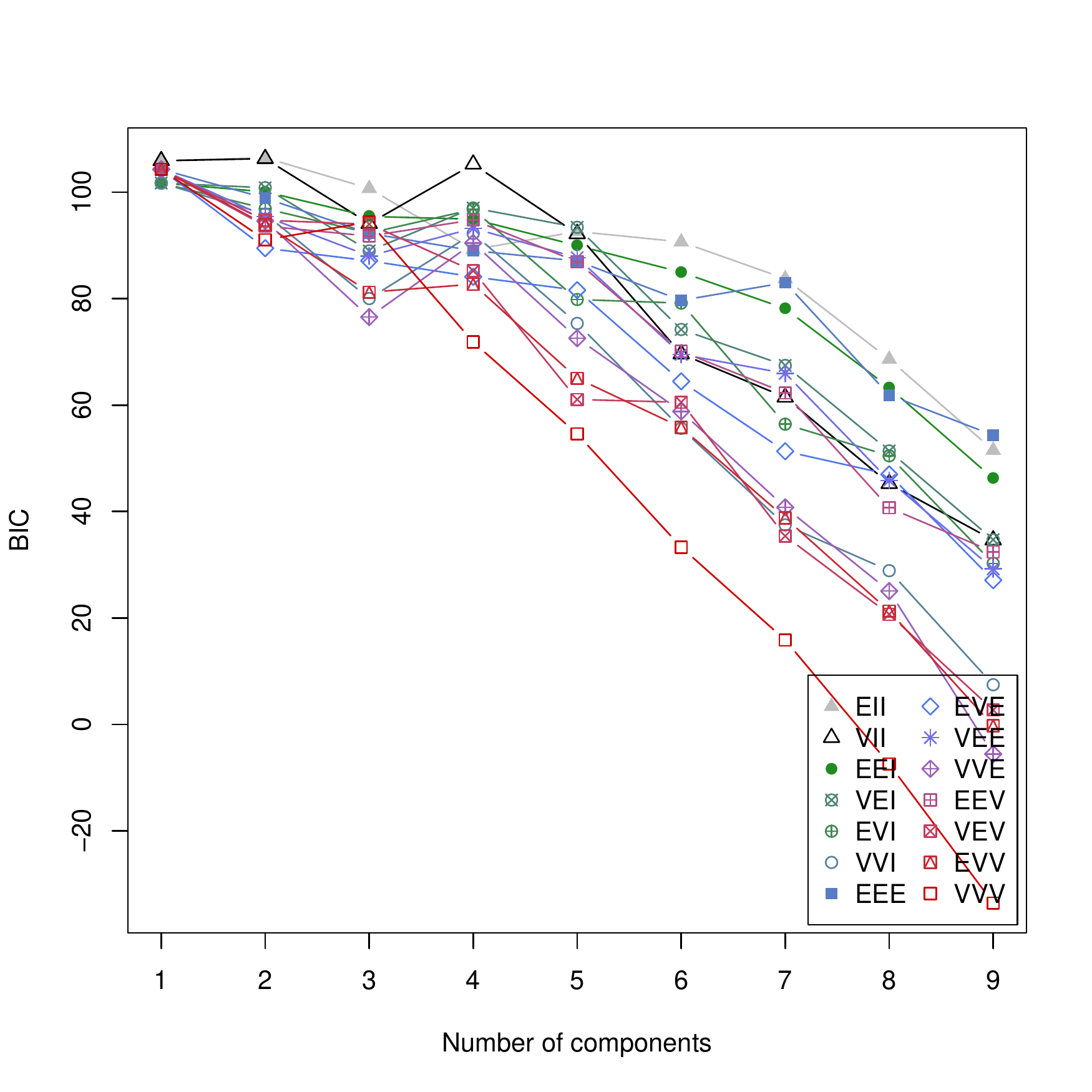}
      \includegraphics[width=8.5cm]{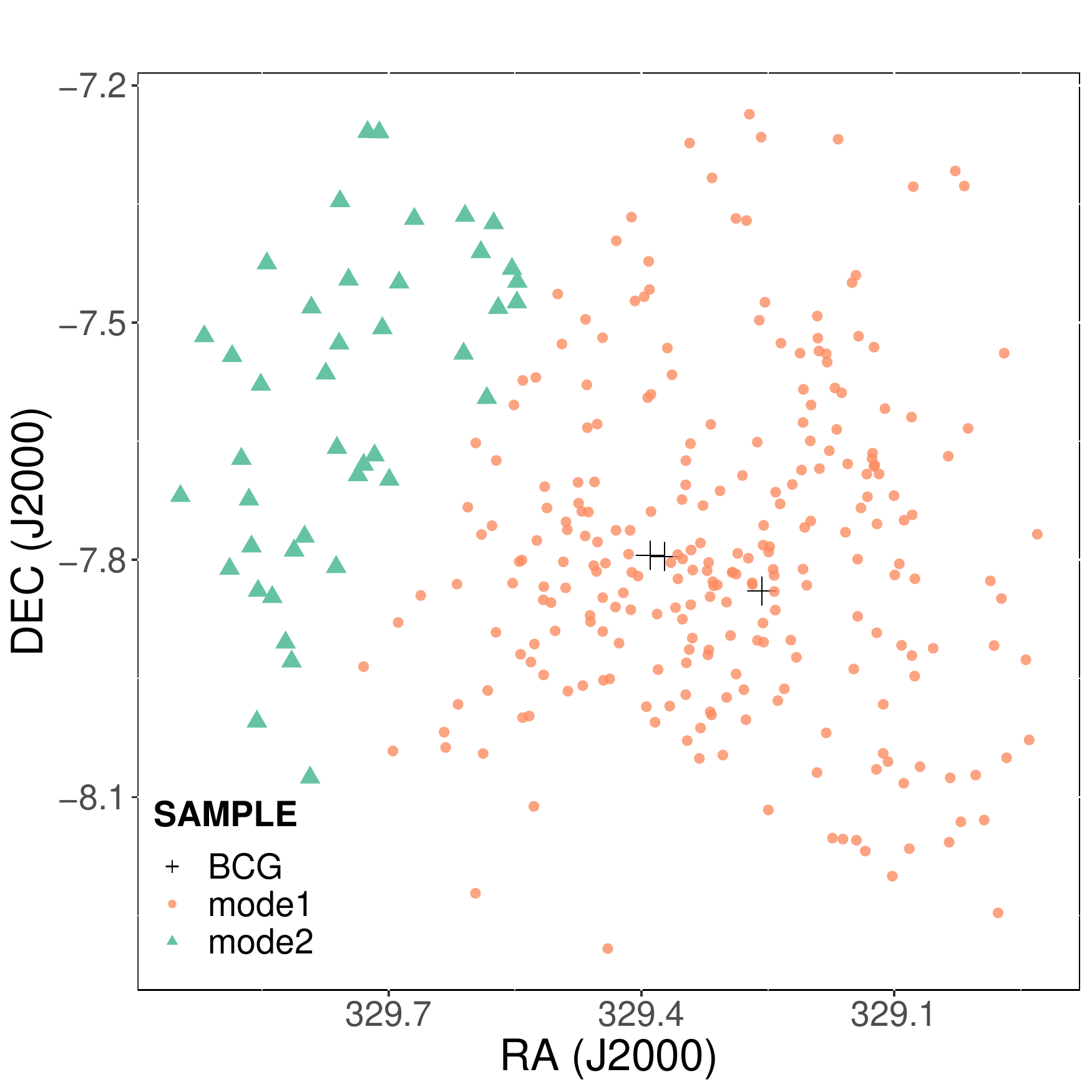}
      \caption{2D analysis. Left panel: BIC analysis for the different models fitted by \textit{mclust}. The two-component model VII was the one that better fitted the data. Right panel: Two subgroups found by the fitting of two-dimensional Gaussians by the GMM. The black crosses mark the positions of the BCGs.}
      \label{fig:2D}
\end{figure*}

    Our 2D analysis found two groups in the clustercentric radius of 2.6\,Mpc. In Fig.~\ref{fig:2D} the left panel shows the variation of the different BICs. The model with the highest BIC is the VII (spherical Gaussians with variable volume and equal shape), that found two subgroups in the sample. The right panel shows the two groups encountered by the fitting of the two-dimensional Gaussians in the plane of the sky. The black crosses mark the position of the BCGs. However, several models fitted by \textit{mclust} have close BICs which makes it difficult to obtain a reliable estimate of the number of subgroups in the sample.

\subsubsection{3D Analysis}
\label{subsec:3D_analysis}
    To study 3D substructures, one can search for over-densities in the galaxy distribution using the two coordinates of the plane of the sky (RA and Dec.) along with the velocity in the line of sight as a proxy for a third spatial coordinate.
    
\begin{figure*}
      \includegraphics[width=8.5cm]{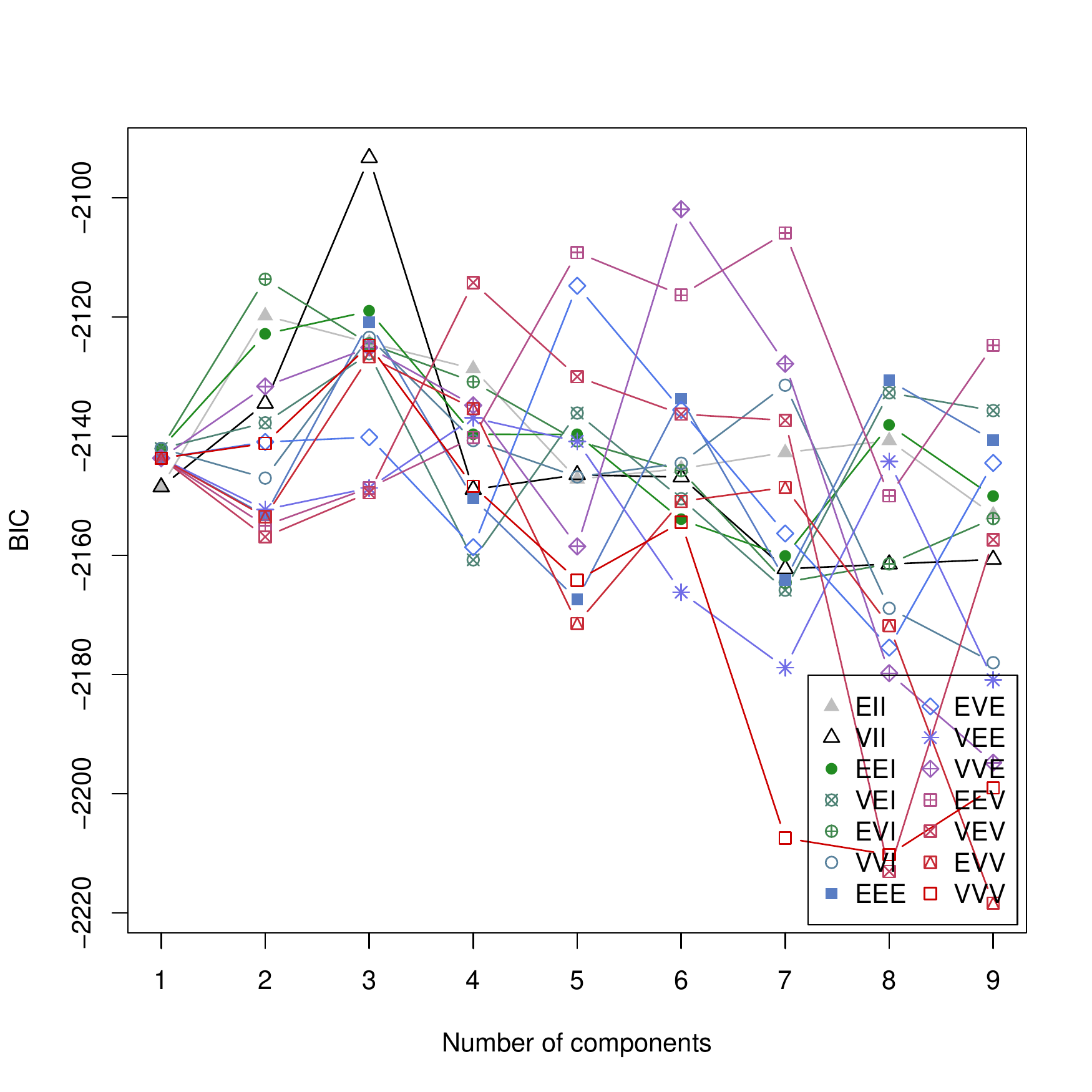}
      \includegraphics[width=9.0cm]{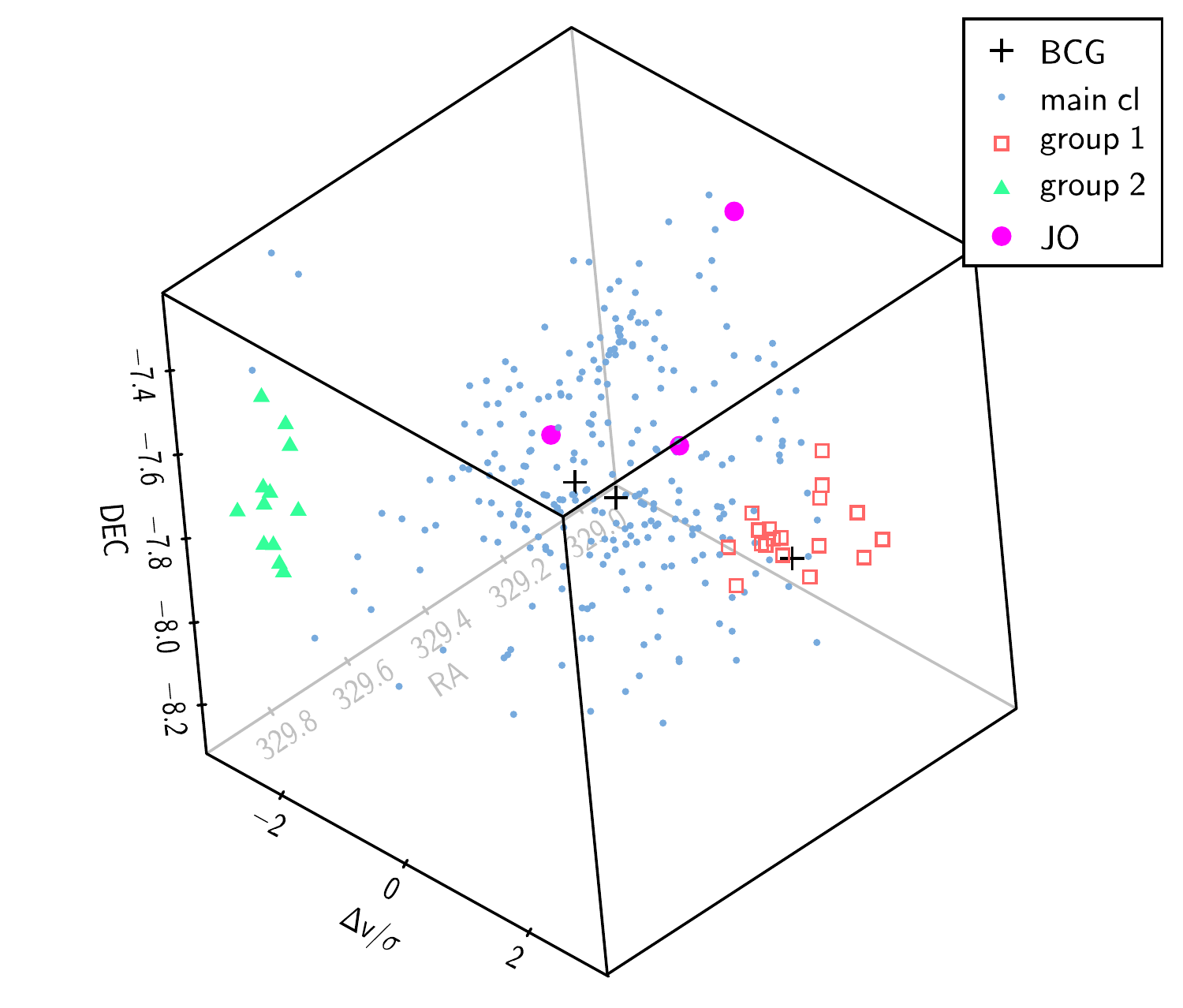}
      \caption{3D analysis. Left panel: BIC analysis for the different models fitted by \textit{mclust}. The three-component model VII is the one to better fit the data. Right panel: Distribution of galaxy groups found by \textit{mclust} in 3D. The axes were scaled in order to improve the fitting. Jellyfish candidates from \protect\cite{Poggianti2016} in A2399. and BCGs are shown as pink filled circles and black crosses, respectively.} 
      \label{fig:3D}
\end{figure*}

\begin{figure*}
      \includegraphics[width=8.5cm]{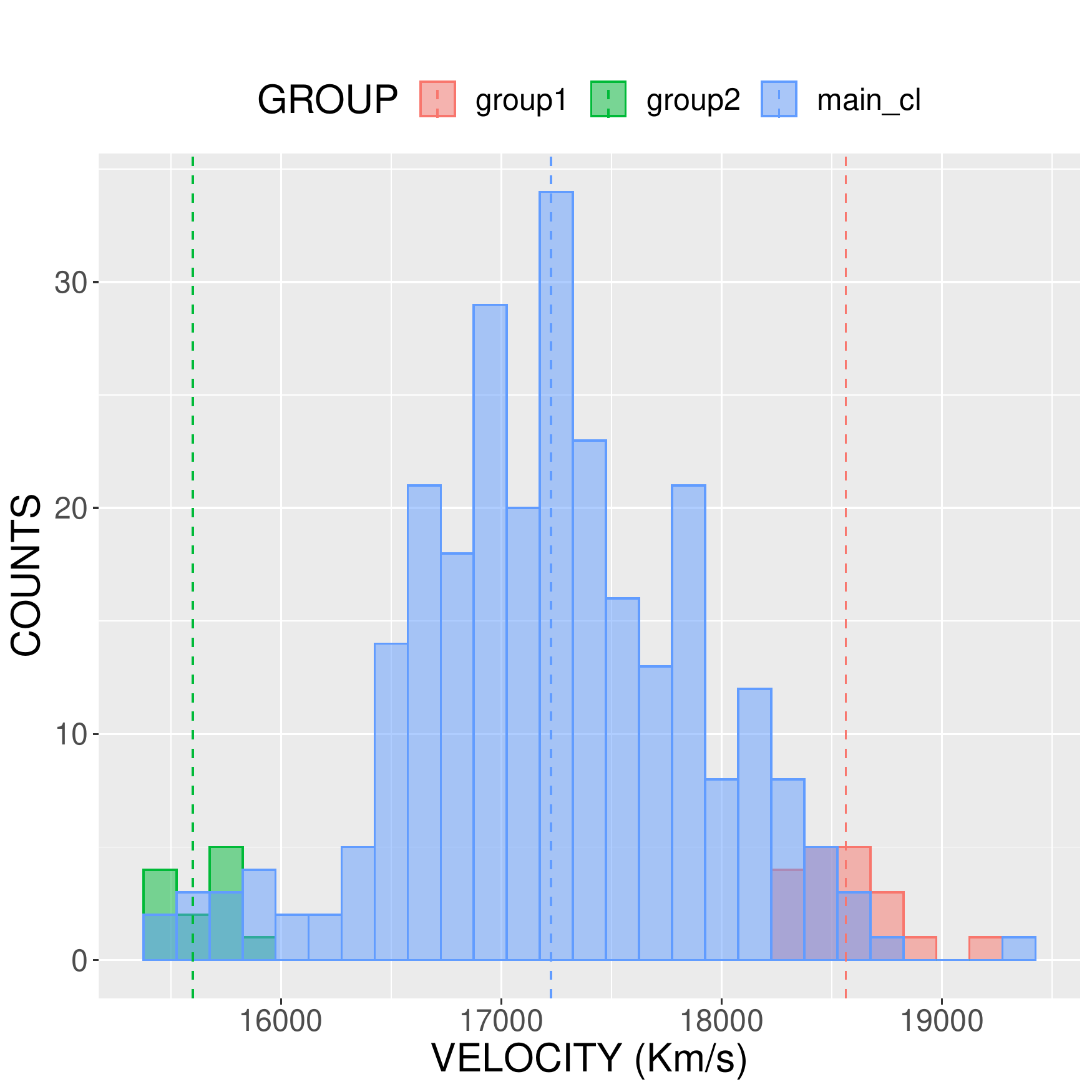}
      \includegraphics[width=9.0cm]{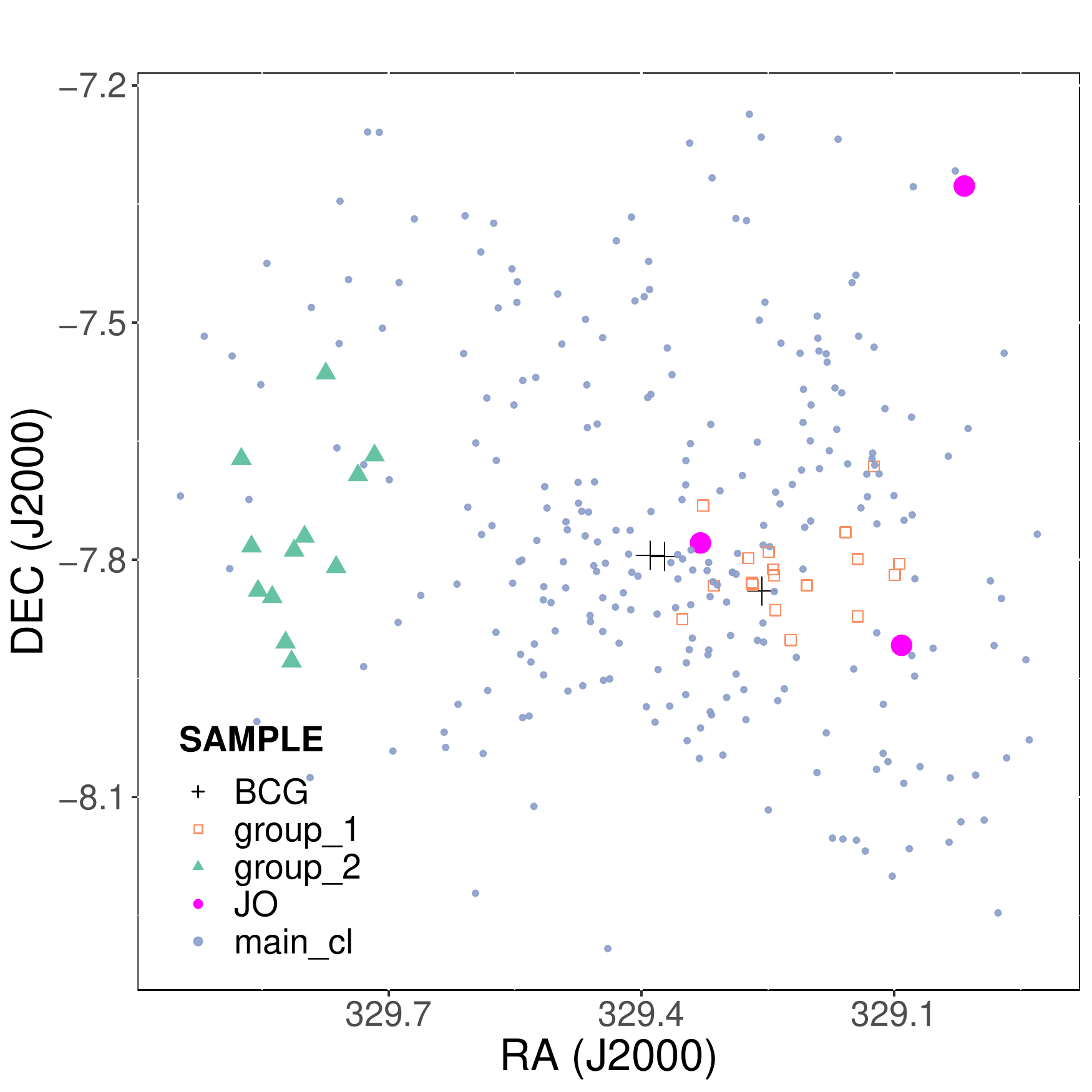}
      \caption{Left panel: Velocity histogram of our sample highlighting the groups found in the 3D optical analysis with different colours. The traced lines indicate the average velocities of each group. Right panel: Plane of the sky distribution of the 3D groups found by \textit{mclust}. The filled pink circles indicate the position of three jellyfish galaxies from \citet{Poggianti2016} in A2399. BCGs are shown as black crosses.} 
      \label{fig:3D_2}
\end{figure*}

\begin{figure}
      \includegraphics[width=8.0cm]{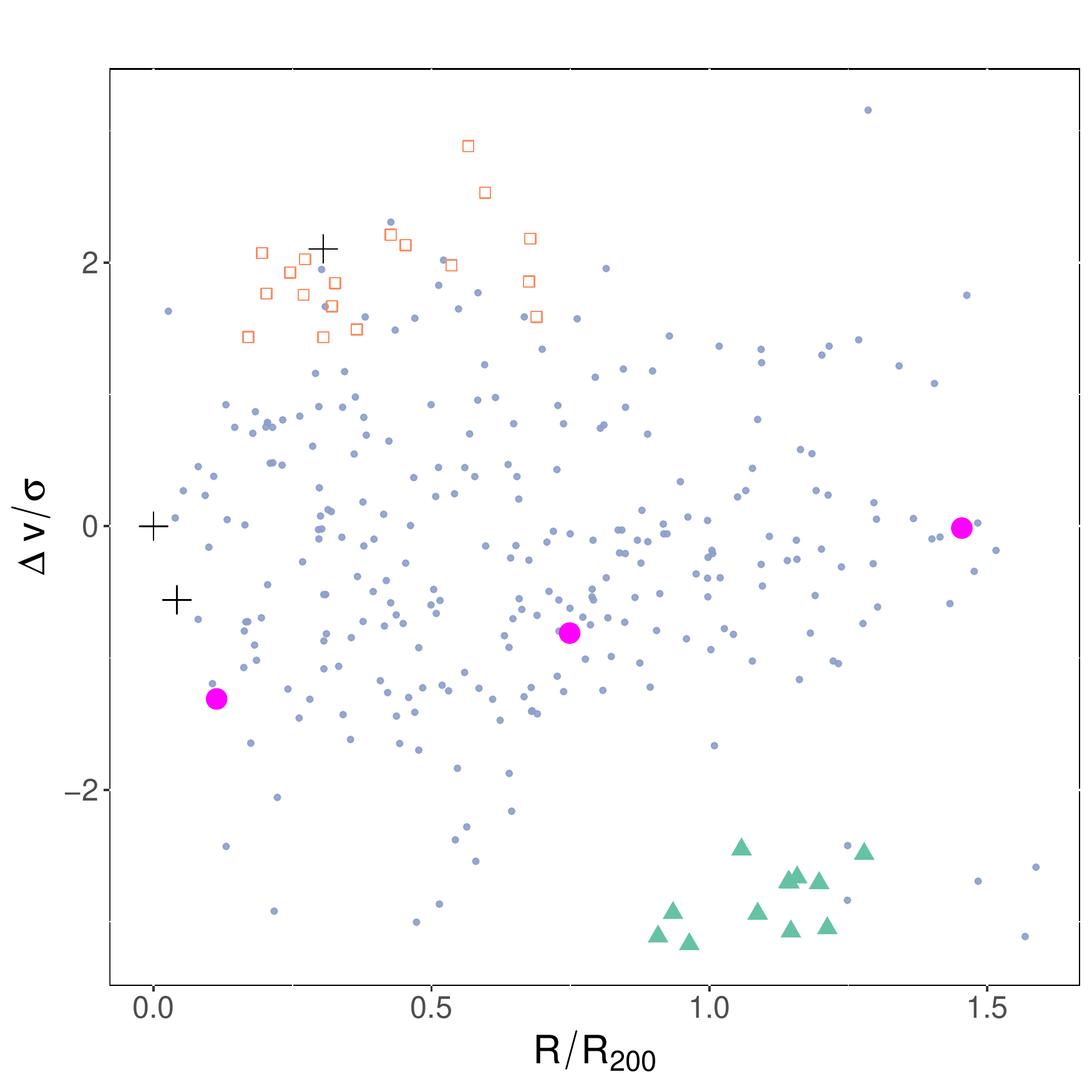}
      \caption{The phase space diagram of A2399 indicating the position of the three jellyfish galaxies from \citet[]{Poggianti2016} (filled pink circles) and the groups found in the 3D analysis by \textit{mclust}. The green filled triangles represent group 2, easily distinguishable in the phase space diagram, while the red open squares (group 1) is very entangled to the main cluster. BCGs are shown as black crosses.}
      \label{fig:phase_space}
\end{figure}

    Our three-dimensional analysis found three groups within the 2.6\,Mpc clustercentric radius. In Fig.~\ref{fig:3D} the left panel shows the BICs of the different models fitted. The three-component model VII was the one selected again. The BIC difference from this three-component model and the second larger BIC is greater than 10. This fitting is, therefore, more reliable than the previous cases \hbox{\citep{Kass1995}}. The right panel shows the distribution of the three groups found by \textit{mclust} in 3D. In the 3D analysis, \textit{mclust} retrieved the main cluster (blue dots) of A2399 and two groups in the infalling region (group 1 shown as open red squares and group two as filled green triangles). BCGs are shown as the black crosses. The pink filled circles indicate the positions of galaxies under ongoing gas stripping, the so-called jellyfish candidates from \cite{Poggianti2016}, that will be discussed later in Sec. \ref{subsec:gas_morph}. The axes were scaled in order to improve the fitting since velocity is given in values that are much bigger than RA and Dec. The left panel of Fig.~\ref{fig:3D_2} shows the velocity histogram highlighting the classification found by \textit{mclust} in 3D. The right panel shows the same groups distributed in the plane of the sky with the same colours. In Fig.~\ref{fig:phase_space} the same groups are plotted in the phase space diagram represented by the same colours as before. In the phase space (PS) diagram it is clear that the green group is separated from the main body of the cluster. The red group, on the other hand, is more difficult to identify in the PS diagram by eye. However, further on, we will see that this group seems to be associated with a clump in the X-ray emission.
	
	In summary, the 1D and 2D analysis found probable interactions in the line of sight and plane of the sky, respectively, but the BIC gaps between the most and the second most probable models were smaller than 5, therefore, their results were not trustful. Fortunately, in 3D the best model found the main cluster and two extra groups. In this case, the difference between the first and second highest BICs was higher than 10, so the result was trustful. For this reason, from now on we will just work on the further investigation of the substructures found by the 3D analysis.
	
\begin{table*}
\renewcommand{\arraystretch}{1.5}
	\centering
	\caption{Main properties of A2399 and its subgroups.}
	\label{tab:cluster_prop}
	\begin{tabular}{lcccccccccc} 
	\hline
	Cluster & RA       &Dec.      & $z_{\rm spec}$ & $ N_{\rm spec}$ & $N_{\rm memb}$ & $M_{200}$            & $R_{200}$ & $M_{500}$            & $R_{500}$ & $\sigma_{p}$  \\
		&(J2000)   &(J2000)  &        &          &           & ($10^{14}\,{\rm M}_{\odot}$) & (Mpc)     & ($10^{14}\,{\rm M}_{\odot}$) & (Mpc)     & (km\,s$^{-1}$)        \\
	\hline
	Parent  &329.37258 &$-$7.79569 &0.0579  &299       &275        &$5.25^{+0.45}_{-0.39}$              &  1.63 & $4.84^{+0.41}_{-0.36}$             & 1.17  & $621.2^{+26.4}_{-23.1}$      \\
	Main cl &329.34457 &$-$7.78231 &0.0575  &268       &256        &$4.25^{+0.40}_{-0.32}$              &  1.52 & $3.95^{+0.37}_{-0.30}$             & 1.10  & $552.4^{+25.9}_{-21.0}$       \\
	Group 1 &329.24268 &$-$7.81965 &0.0619  &19        &18         &$0.22^{+0.13}_{-0.07}$              &  0.56 & $0.20^{+0.12}_{-0.06}$             & 0.41  & $216.2^{+62.1}_{-32.7}$       \\
	Group 2 &329.81361 &$-$7.78557 &0.0521  &12        &12         &$0.12^{+0.06}_{-0.03}$              &  0.47 & $0.12^{+0.06}_{-0.03}$             & 0.34  & $146.3^{+36.8}_{-13.6}$       \\
	\hline
	\end{tabular}
\end{table*}

\subsection{Physical Parameters of the Optical Substructures}

    The \textit{mclust} 3D analysis described above provided the centroids and redshift of each structure identified (i.e. the main cluster and two groups). We then used the \textit{shifting gapper} technique to select only the members of each group. The procedure used to estimate the velocity dispersion $\sigma_{P}$, characteristic radii ($R_{500}$ and $R_{200}$), and masses ($M_{500}$ and $M_{200}$) was the same as \citet{Lopes2009}. First, we calculate the robust velocity dispersion for each group using the biweight estimator \citep{Beers1990}. Next, we derive the projected virial radius and then the virial mass is calculated following \citet{Girardi1998}. $R_{500}$, $R_{200}$, $M_{500}$ and $M_{200}$ were estimated assuming an NFW profile \citep*{Navarro1997} as in \citet{Lopes2009}.
	
	Table~\ref{tab:cluster_prop} lists the main properties we obtained for the parent cluster and its subgroups within $\sim$2.6\,Mpc. The first column identifies each analysed group. The second, third, and fourth columns contain the centroids of the coordinates (RA, Dec.), and redshift ($z_{\rm spec}$), respectively. The fifth and sixth columns list the number of galaxies with spectroscopy in each subgroup, $N_{\rm spec}$, and the number of members selected by the \textit{shifting gapper}, $N_{\rm memb}$. Columns seven, eight, nine and ten list the characteristic radii and masses of the subgroups $M_{200}$, $R_{200}$, $M_{500}$, and $R_{500}$. The last column lists the velocity dispersion. The listed values indicate the main cluster is more than ten times more massive than the two smaller subgroups.

\subsection{Search for interaction in the X-rays}

\subsubsection{Gas morphology and BCGs}
\label{subsec:gas_morph}

    Another way to investigate the dynamical state of galaxy clusters is to inspect how spherically symmetric the distribution of the intracluster gas is.
    
\begin{figure*}
      \includegraphics[width=17.4cm]{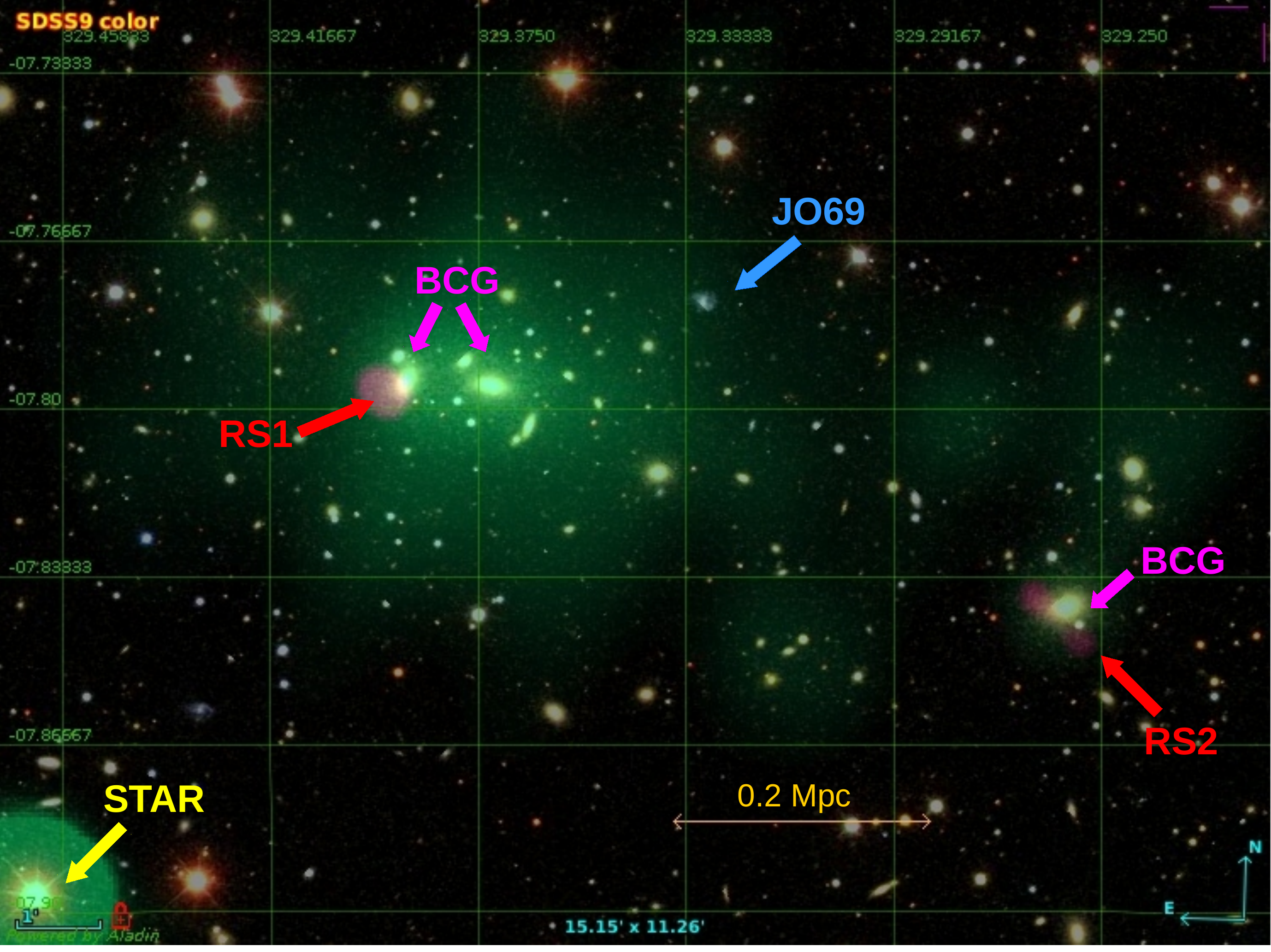}
      \caption{X-ray emission in green and radio emission in red overlaid on the SDSS optical image of A2399. Three BCGs are pointed out by pink arrows and one jellyfish galaxy pointed by a cyan arrow, JO69. This image corresponds to a very central region of A2399 within a radius of 0.5\,Mpc.}
   \label{fig:multiwavelength}
\end{figure*}

    Fig.~\ref{fig:multiwavelength} shows that the X-ray emission for A2399 (in green) is elongated to the southwest-northeast direction, suggesting an interaction scenario in the plane of the sky, such as in the optical 3D analyses. The X-ray emission has a main concentration and a clump located in the west. The radio intensity maps of the Very Large Array (VLA) public data are shown as a pink emission. The SDSS coloured image is overlaid on the radio and X-ray ones. Near the centre are two X-ray emission peaks separated by 60\,arcsec ($\sim$68.18\,kpc) on the plane of the sky. The central BCG, SDSS J215729.42-074744.5, is at the same location as the first X-ray peak and does not emit in radio frequency. The other BCG at the centre, RS1 radio source, is the radio galaxy SDSS J215733.47-074739.2 which coincides with the second X-ray peak in the central region. In the southern part of the X-ray western clump, we can see a third BCG that is also a radio source, RS2. This galaxy is SDSS J215701.72-075022.5. The bright X-ray source in the lower right corner corresponds to a foreground star. We also highlight JO69, a jellyfish candidate from \cite{Poggianti2016} at RA $= 329.\!\!^{\circ}36517$, and Dec. $= -7.\!\!^{\circ}80358$, with a cyan open circle. The other two jellyfish candidates in this cluster fall outside of the X-ray emission. In the next subsection, we will discuss the location of this jellyfish with respect to the spectral maps.   

\begin{figure*}
      \includegraphics[width=8.7cm]{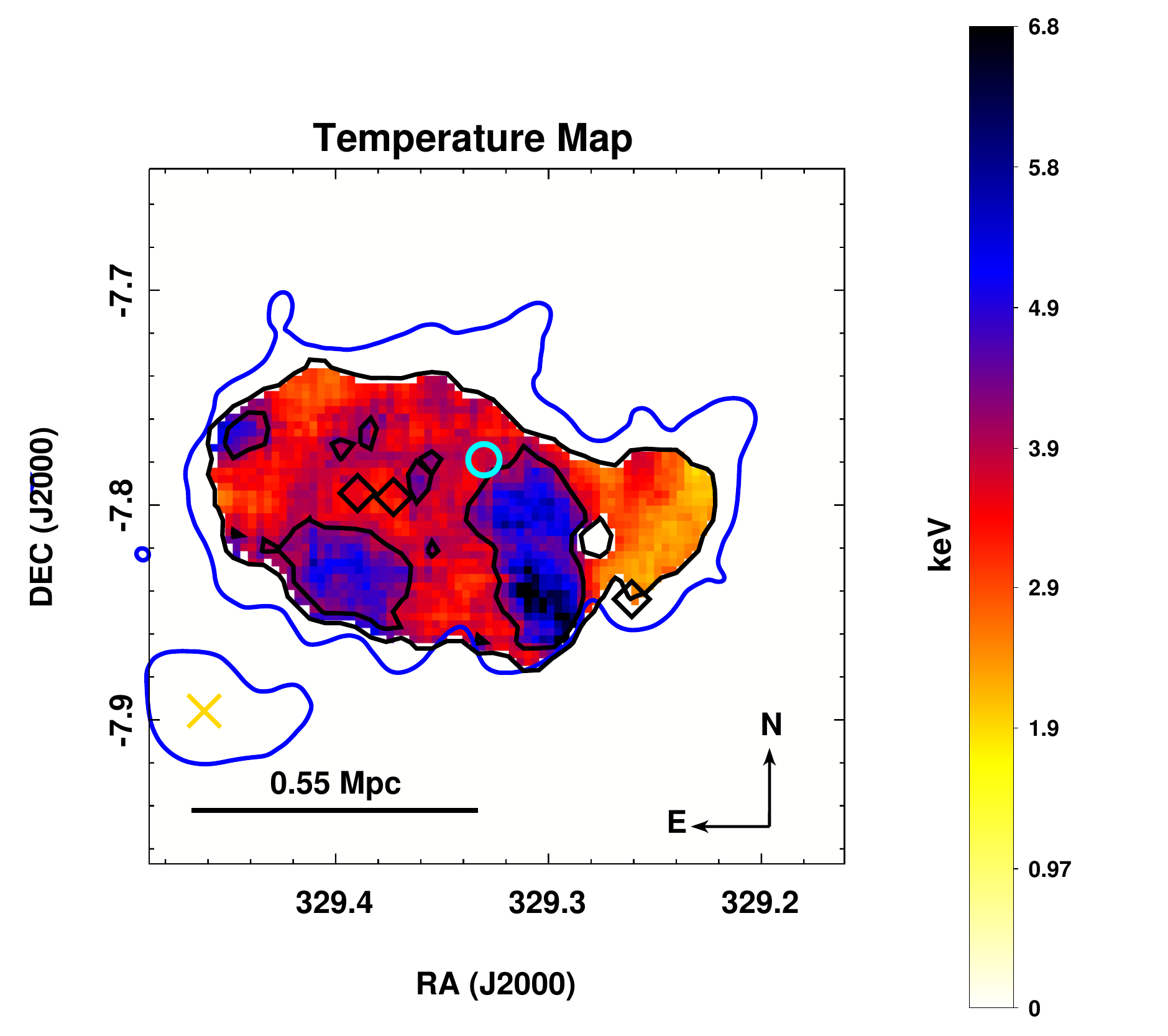}
      \includegraphics[width=8.7cm]{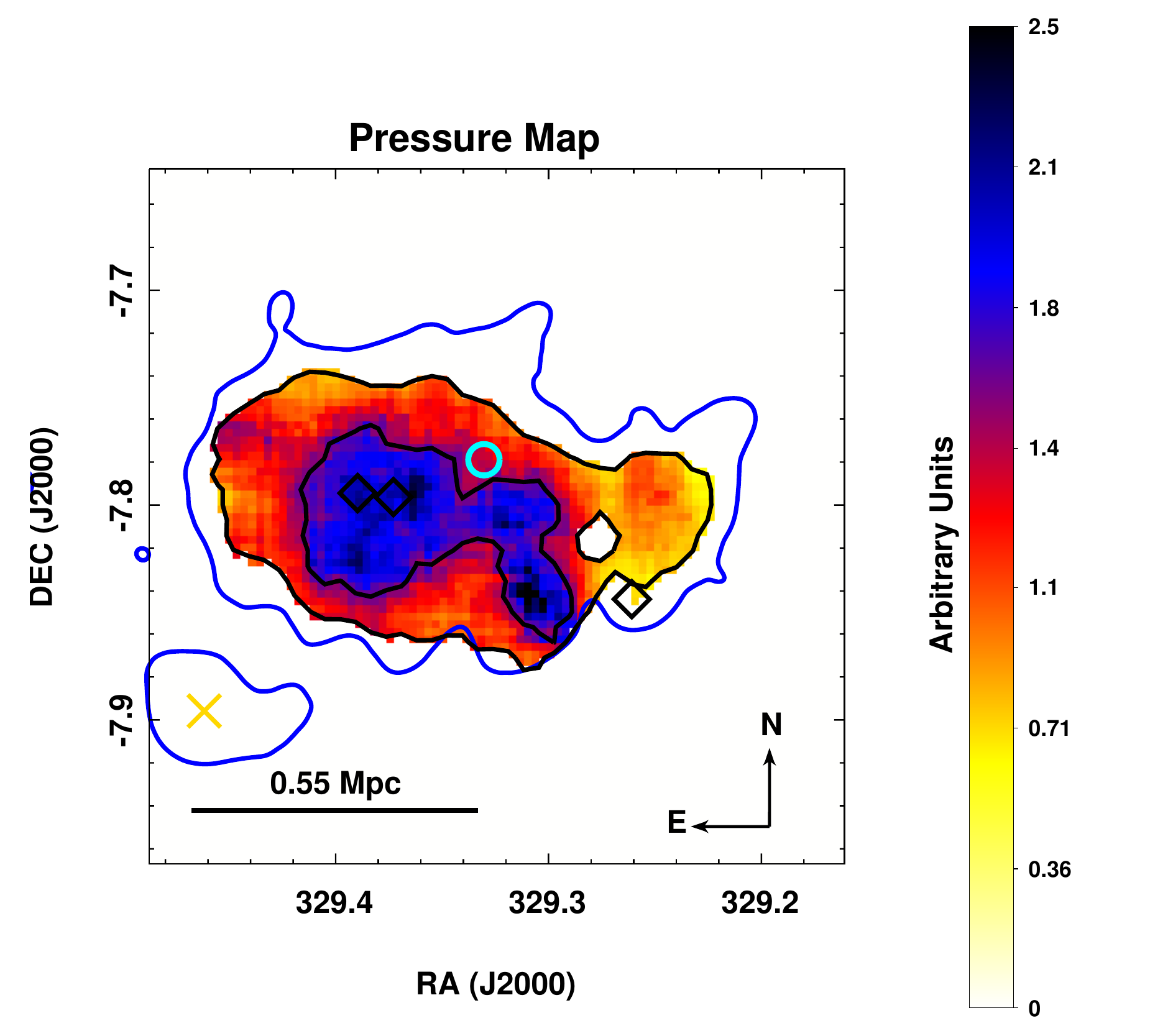}
      \includegraphics[width=8.7cm]{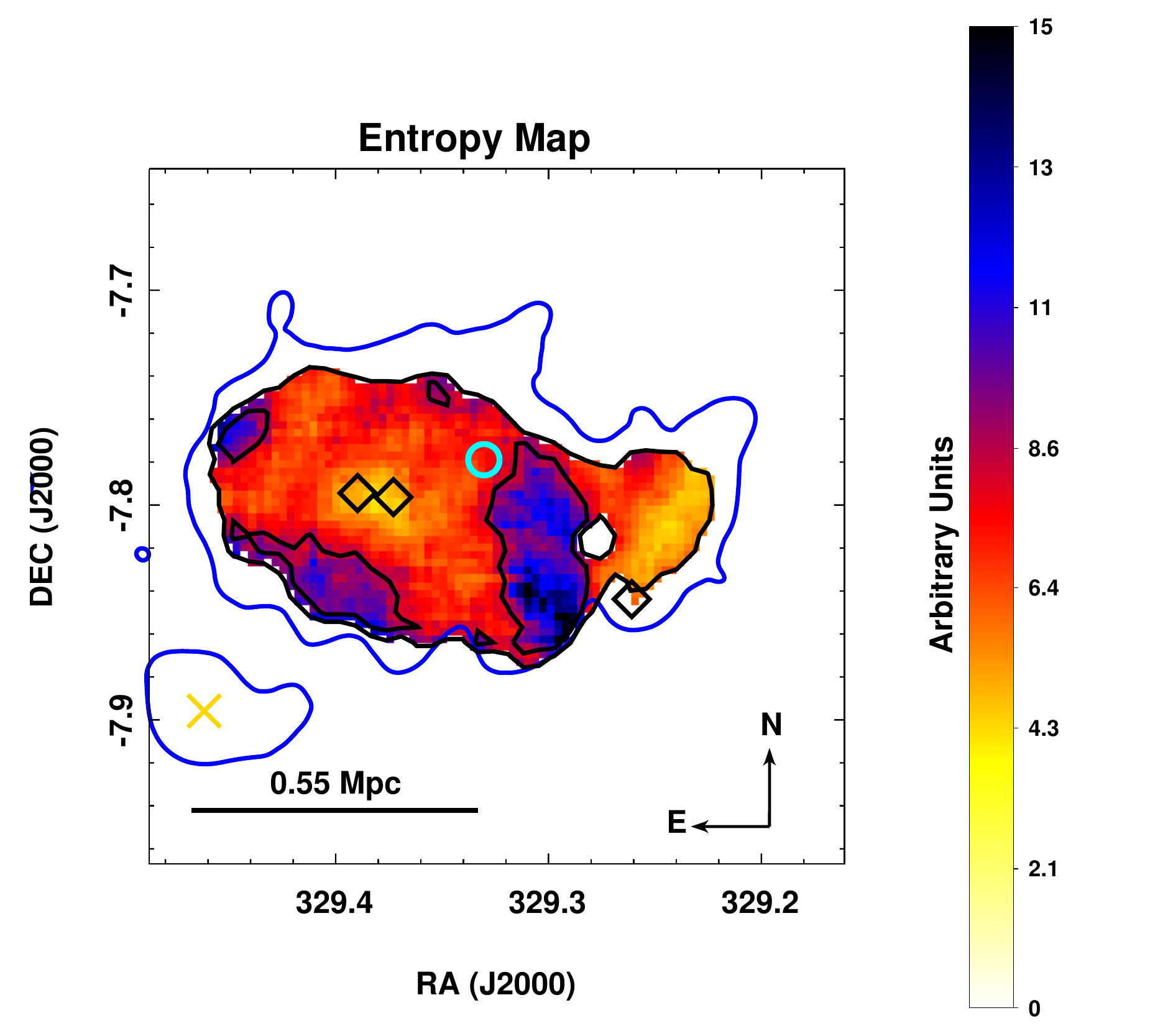}
      \includegraphics[width=8.7cm]{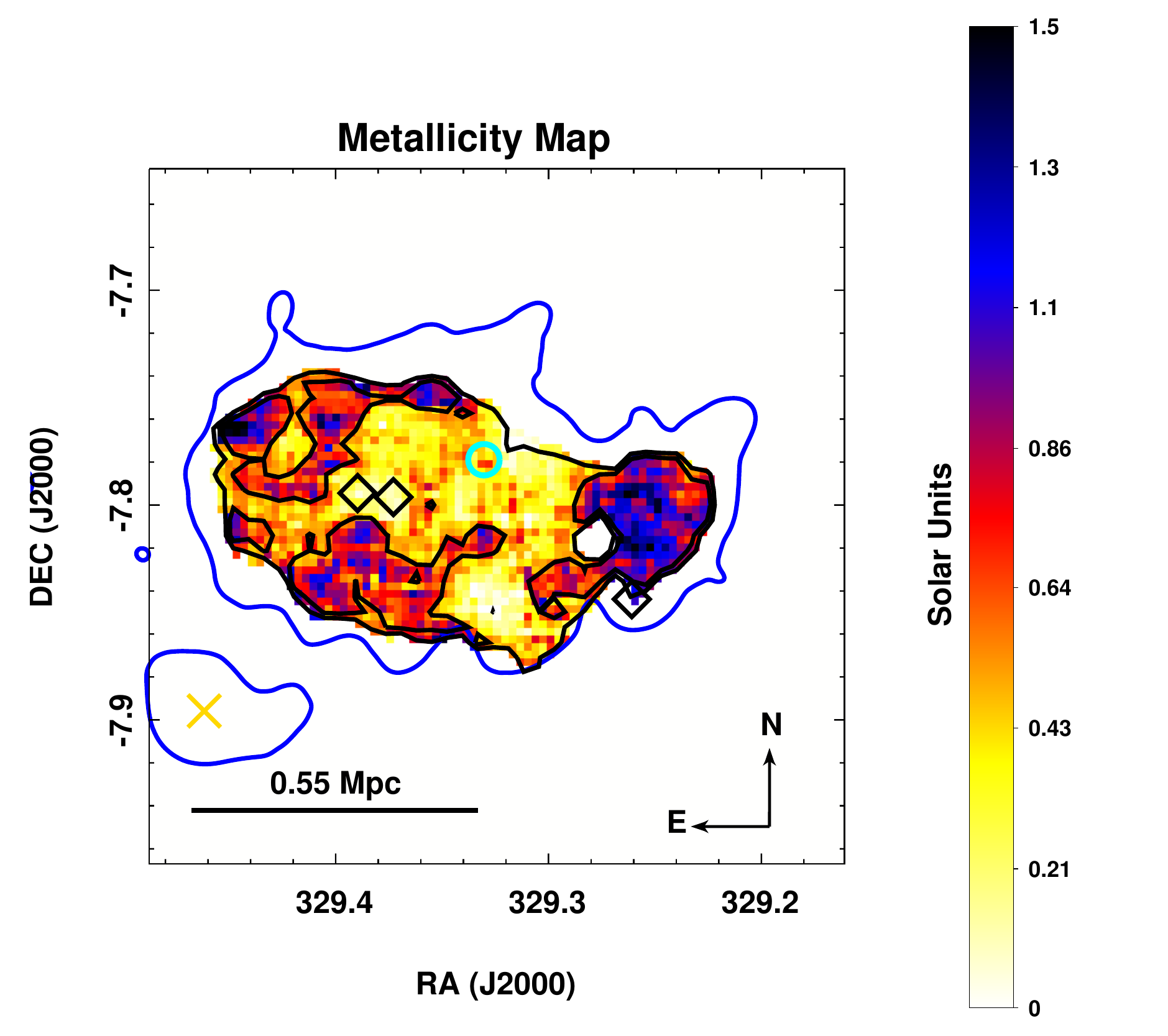}
      \caption{X-ray spectral maps. The open circles in cyan show the position of JO69, a jellyfish galaxy. The open black diamonds indicate where the BCGs are located, and the yellow crux marks the position of a foreground star. The blue contour shows the X-ray emission from \textit{XMM-Newton}. All the maps show that A2399 is a disturbed cluster with a western clump that resembles a bullet.}
      \label{fig:maps}
\end{figure*}

\subsubsection{Structure of Thermodynamical Maps} 

    Spectral maps are a powerful tool for understanding the dynamical state of a galaxy cluster. Interactions between clusters can leave different imprints in the gas, such as discontinuities and asymmetries in the temperature, pressure, and entropy maps.
    
    Fig.~\ref{fig:maps} shows the thermodynamic (temperature, pressure, and entropy) and metallicity maps derived from the \textit{XMM-Newton} observation. Overall, the maps have a complex structure. There is a western clump with lower temperature ($\sim$ 2--3\,keV), pressure, and entropy, although higher metallicity. Also, we can notice a region extended in the north-south direction, to the left of the western clump, with higher temperature (about 6\,keV) and spatially coincident with a high-entropy region. In the pressure map, this region has also higher pressure values, although the region is more elongated, suggestive of a shock region. Near this discontinuity, at the northeast, is the jellyfish candidate JO69, shown as a cyan open circle, which probably had its morphology modified by the enhancement of the winds near to this shock region. In \citet{Owers2012} and in \citet{Vijayaraghavan2013}, the authors found a correlation between the ram pressure efficiency and the existence of shocks in the ICM. In fact, two more jellyfish candidates (JO68 and JO70) from \citet{Poggianti2016} were identified in A2399 at larger clustercentric radii.
    
    In recent work, \citet{Mitsuishi2018} have found that A2399 has very high entropy in the centre for a low surface brightness cluster. The authors concluded that the AGN, near to the centre, cannot be responsible for the high entropy in the central region of this cluster. The authors proposed that a merger event caused such an increase. In our pseudo-entropy map, we have regions of higher entropy out of the core that are suggestive of merging as well.

\subsubsection{Optical and X-ray comparison}

\begin{figure*}

	\includegraphics[width=17.0cm]{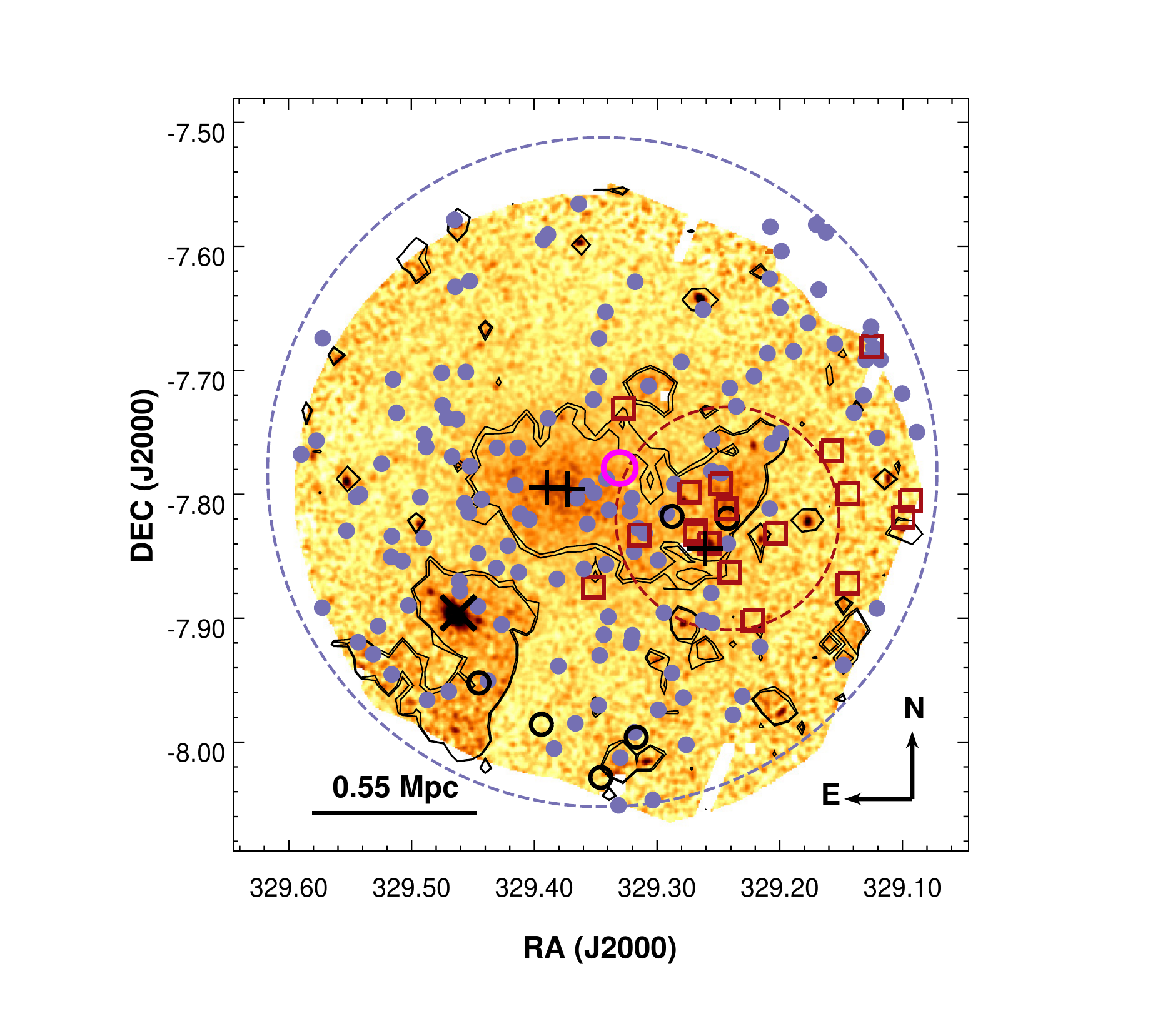}
	\caption{Mclust groups in the central region of A2399 overlaid on the ICM. The filled blue circles represent the main cluster and the open red squares the group 1. Black contours highlight the X-ray emission and black crosses show the positions of the BCGs. Interlopers are marked by black open circles. The dashed blue and red circles show $R_{500}$ of the main cluster and group 1 respectively. The open magenta circle shows JO69.}
    \label{fig:opt_groups_xray}
\end{figure*}

    In Fig.~\ref{fig:opt_groups_xray}, we show the main cluster and group 1 found in the 3D analysis of the \textit{mclust} as blue filled circles and red open squares, respectively. The optical groups are plotted on the top of the merged images of the XMM-Newton EPIC cameras. Group 2 falls outside the X-ray image. The $R_{500}$ of each group is shown as a dashed circle in the same colour as the corresponding groups. The X-ray emission is highlighted by the black contours and the jellyfish JO69 is shown as a pink open circle. The image suggests that group 1 is associated with the western extension of the X-ray emission. Analysing this image together with the spectral maps we suspected that A2399 is a bullet cluster-like, where group 1 crossed the main cluster generating the characteristics observed in the thermodynamic maps. To test this hypothesis we performed a series of hydrodynamical simulations that will be shown in the following section.
    
    It is worth mentioning that \textit{mclust} was successful in finding groups associated with the emission of the main cluster and the western clump in X-ray. If this trend is confirmed in a larger sample, the identification of bullet-like clusters can be automated, increasing the sample of known clusters of this type. This would be of fundamental importance in restricting dark matter parameters such as its cross-section \citep{Markevitch_2002}.

\section{Hydrodynamical Simulation}
\label{sec:hid_sim}

In order to evaluate whether A2399 could be the result of a recent collision, we performed a set of hydrodynamical $N$-body simulations. Here we report on one model that reproduced some of the observed features. The scenario we wish to evaluate is that the clump in the west may have passed through (or near) the core of the main cluster. The main observational constraints the simulation aims to fulfil are the temperature structure (i.e.~the presumed shock front with 5.2--6.3\,keV); and the current separation between the centroids of the two objects. We assume that the separation between the simulated dark matter peaks in the simulation is a proxy for the observed separation between the BCG of the main cluster and the BCG of the western clump.

The simulation setup consists of two idealised galaxy clusters that are initially spherical and relaxed. They are set to collide with a given impact parameter and a given initial velocity. Each galaxy cluster is modelled by gas particles and dark matter particles. In the main cluster, the gas and the dark matter follow a \cite{Hernquist1990} profile. The parameters were chosen such that its virial mass is $M_{200} = 4.0 \times 10^{14}\,{\rm M_{\odot}}$ and its virial radius is $R_{200}=1.5$\,Mpc. The choice of density profiles fixes the gas temperature profile through the imposition of hydrostatic equilibrium. The temperature around $R\sim$100\,kpc is $\sim$3.8\,keV but there is a small cool-core. In the western clump, a \cite{Hernquist1990} profile was used for the dark matter and a \cite{Dehnen1993} profile was adopted for the gas. The choice of parameters was such as to give $M_{200} = 0.2 \times 10^{14}\,{\rm M_{\odot}}$ and $R_{200}=0.6$\,Mpc, which results in a central gas temperature of $\sim$2\,keV. The initial condition parameters were chosen to be close to the observational estimates from the optical and X-ray. The numerical realisations of these initial conditions were generated with the techniques described in \cite{Machado2013}. In the model presented here, each cluster has $10^{6}$ dark matter particles and $10^{6}$ gas particles. After the two clusters have been created, the collision is prepared in the following way. The clusters are positioned with an initial separation of 3\,Mpc along the $x$ axis, and also with an impact parameter $b$ along the $y$ axis. The initial velocity between them is $v_0$, parallel to the $x$ axis. This impact parameter $b$ refers to the $y$ offset at $t=0$; later on, at the moment of pericentric passage, the separation is generally much smaller. We found that the initial velocity of $v_0=-700$\,km\,s$^{-1}$ produced suitable results. The choice of this approximate value was motivated by the result of the two-body dynamical modelling, which provides an estimate of the total mechanical energy of the system. For simplicity, we assume that the collision takes place on the plane of the sky.

The simulations were performed with the hydrodynamical $N$-body code \textsc{Gadget-2} \citep{Springel2005}. The evolution of the system was followed for 5\,Gyr but the relevant phases take place within the first 3\,Gyr. Cosmological expansion is ignored due to the relatively small spatial extent of the problem.

One desirable feature of the simulated temperature map is that the presumed shock front (blue region in Fig.~\ref{fig:maps}a) needs to be roughly perpendicular to the line connecting the BCGs. This configuration would arise naturally in the case of a head-on encounter, in which the velocity of the clump is entirely radial towards the main cluster and thus the shock front is perpendicular to it. However, in such a perfectly frontal collision ($b=0$), the outcome of the simulation would be symmetrical to a degree that is not warranted by the complex observed morphologies. At the other extreme, if the impact parameter is too large, the pericentric distance may be considerable. In this regime of large $b$, at the moment when the western clump overtakes the main cluster, its velocity is mostly perpendicular to the line connecting them. Therefore, the shock front would be approximately parallel to the line connecting the two clusters, which is not the desired result. Therefore, a suitable compromise must lie in a choice of $b$ that leads to a non-zero but small pericentric separation. By means of a limited but plausible exploration of the parameter space, we found that an initial impact parameter of $b=500$\,kpc gives satisfactory results. At the moment of pericentric passage, the separation between the centroids of the clusters is 110\,kpc. This may seem small, but it is known that even such modest pericentric distances are capable of bringing about substantial departures in comparison to perfectly frontal collisions \cite[e.g.][]{Machado2015,Doubrawa2020}. Choices of $b$ that led to pericentric distances greater than about 200\,kpc gave poorer results. 

\begin{figure}
\includegraphics[width=\columnwidth]{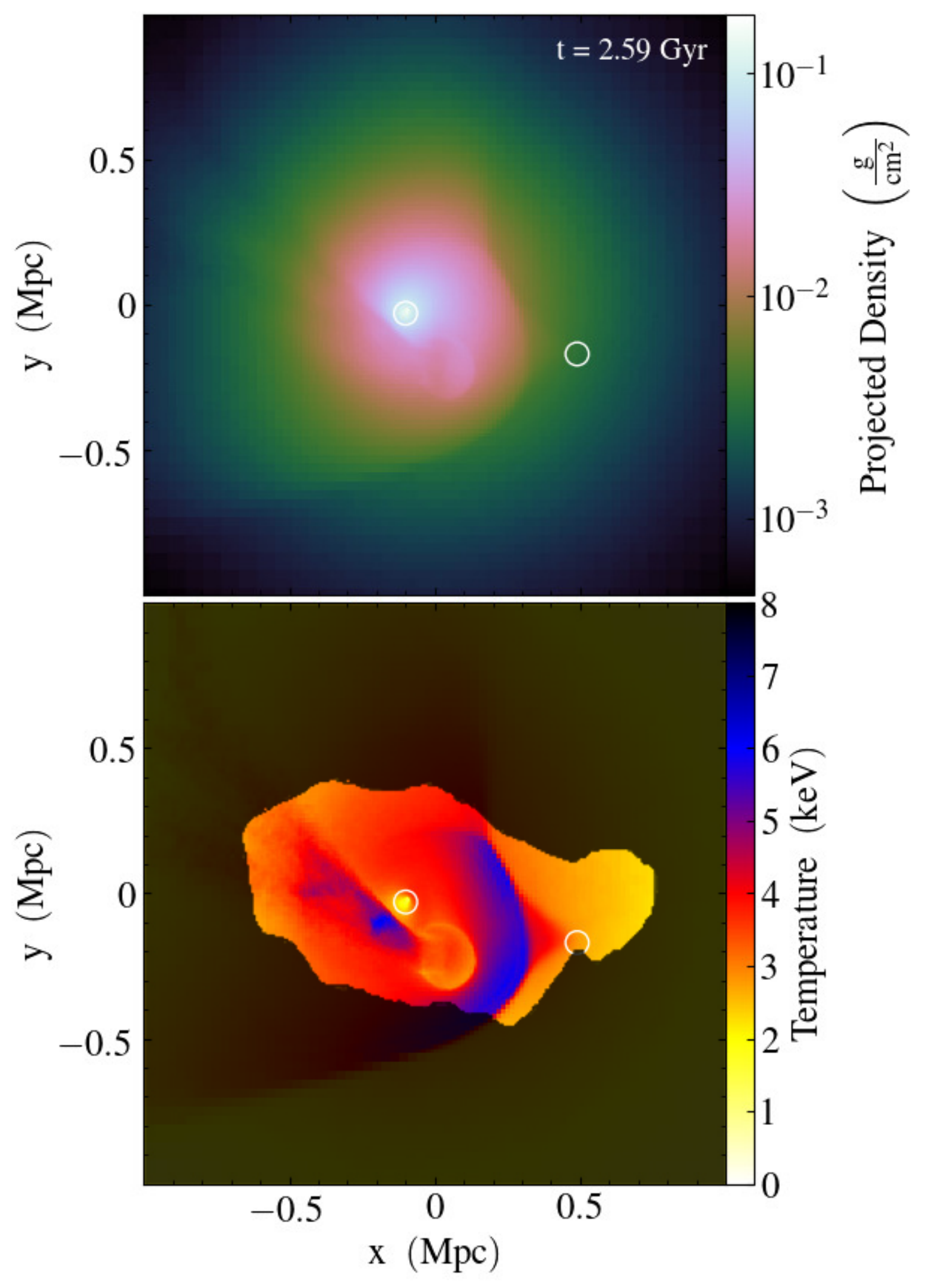}
\caption[]{The best-matching instant of the simulation. Top: projected density map. Bottom: emission-weighted projected temperature map; the overlaid mask outlines a region comparable to the region where observational data are available (Fig.~\ref{fig:maps}). The white circles mark the positions of the dark matter peaks of each cluster.}
\label{sim1}
\end{figure}

In our best simulation model, the pericentric passage takes place at $t=2.35$\,Gyr. Approximately $0.24$\,Gyr later, a configuration is reached when the simulated temperature map best reproduces the main features of the observed temperature map. Simultaneously, the separation between the dark matter peaks reaches 580\,kpc -- within an error of five per cent of the observed separation of 550\,kpc. Therefore we regard this instant ($t=2.59$\,Gyr) as the best-matching instant of the simulation. The density map and the emission-weighted temperature map of the best instant are shown in Fig.~\ref{sim1}. These projected maps were rotated by an angle on the $xy$ plane merely to match the position angle of the observational images, making the visual comparison more straightforward. There is no inclination between the orbital plane and the plane of the sky.

The temperature structure of the simulated map in Fig.~\ref{sim1} should be compared to the observed temperature map in Fig.~\ref{fig:maps}a. Notice that the region for which observational data are available is restricted, in comparison to the entire simulated frame. In order to produce a fair comparison, a semi-transparent mask was applied to the simulated temperature map in Fig.~\ref{sim1}, outlining a region with the same size and shape as the region where observational data are available. One should also bear in mind that this type of simulation is highly idealised. Nevertheless, this simple simulation succeeds in capturing some of the overall morphological features of the temperature map. There is a hotter region of $\sim$6\,keV, elongated roughly along the north-south direction and which is flanked on both sides by cooler gas; the gas on the side of the main cluster has on average $\sim$4\,keV, and that the gas on the side of the western clump has $\sim$2--3\,keV.

\begin{figure*}
\includegraphics[width=\textwidth]{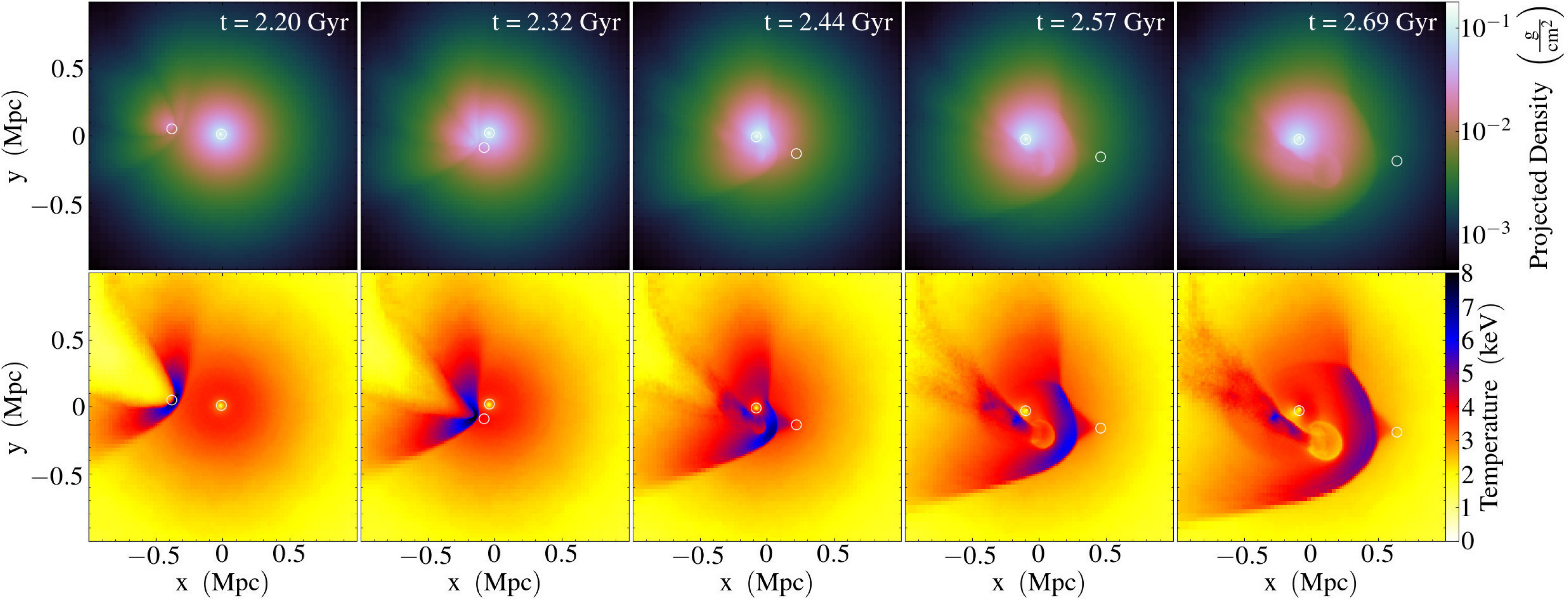}
\caption[]{Time evolution of the simulation. Upper panels show projected density maps and lower panels show emission-weighted projected temperature maps. The white circles mark the positions of the dark matter peaks of each cluster.}
\label{sim2}
\end{figure*}

One of the major shortcomings of idealised binary collisions is the lack of cosmological context and therefore of realistic substructures taking part in the previous history of the clusters. The assumption that the two clusters were initially in hydrostatic equilibrium is adopted for simplicity and also due to lack of prior knowledge of the state of the system. In particular, A2399 could have undergone an earlier collision with the eastern substructure (group 2), perhaps pre-mixing the metals as it passed through the core. This is not modelled by our current simulation. Including more than two structures is not often attempted when modelling individual clusters, because it greatly increases the possible combinations of parameters; a few instances exist \cite[e.g.][]{Bruggen2012, Ruggiero2019, Doubrawa2020}, but they remain rare. Furthermore, in the particular case of A2399, the problem would be even more underconstrained due to the lack of X-ray data at the position of the eastern structure. For these reasons, it would be unreasonable to demand that such simplified simulations reproduce the detailed small-scale structure of a given cluster. Nevertheless, it is interesting to note that the simulation exhibits a plume of hot gas behind the shock front, in a position which is comparable to a similar region of hot gas in the observations. While we cannot claim that they correspond to the same physical process, we can attempt to describe how that feature arose in the simulation. As the shock front passes close to the main cluster core, the nose of the shock is able to advance, but the gas in the western edge of the shock front encounters the denser part of the main cluster core and its motion is hindered, leaving behind a detached wisp of hot gas which will later dissipate. This can be seen in the time evolution of the temperature maps shown in Fig.~\ref{sim2}.

While the simulation succeeds in recovering some global properties of the cluster, it does not reproduce all morphological features in great detail. For example, a more exhaustive exploration of the parameter space of initial conditions might have resulted in an improved model, in which the western clump would have retained more of its gas in the final configuration. Such fine-tuning is beyond the scope of this work. By offering one possible solution, we argue that this type of collision is a plausible scenario to explain the current state of A2399.

\section{DISCUSSION: ON THE IMPLICATION OF THE METALLICITY DISTRIBUTION.}

The ICM is not only composed by primordial elements such as hydrogen and helium but also it is enriched with heavier elements such as iron, which indicates that material processed inside stars is related to this enrichment. There are several mechanisms that can contribute to driving metals from galaxies into the ICM \citep[see][for a review]{Schindler2008}.

Although a strong enhancement in the abundance is found in the central regions of the cool-core clusters \citep{De_Grandi_2001}, \citet{Lagana_2019} showed that several metallicity peaks are observed also out of the core in the metallicity maps of many clusters. Such peaks may be related to galaxies that recently introduced metals into the ICM by galactic winds or RPS but still did not have time to mix with the rest of the gas \citep{Kapferer_2007}.

A plausible hypothesis for explaining the excess of metals in the western clump, that may also account for this `lack' of metals in the centre of the main cluster, is that a possible frontal collision of the western clump with the main cluster enhanced the ram-pressure stripping (RPS) in that region causing the galaxies of group 1 to insert more gas and metals in there. Some hydrodynamical simulations support the idea that collisions increase the RPS \citep{Owers2012,Vijayaraghavan2013}. In fact, in clusters at low redshift like A2399 where the ICM is already formed and the galaxies have already had their star formation peaks around redshift one \citep{Madau_1998}, ram-pressure is the most important mechanism of gas enrichment \citep{Schindler2008}. To test this hypothesis we would need to obtain from the simulations the increase in RPS in the galaxies of the western clump. This will be the subject of future studies.

Usually, metallicity and temperature are anti-correlated \citep{De_Grandi_2001}.
However, we can observe a region in the southeast, and another small region in the northeast where temperatures are higher and metallicities too. Such as in the case presented in \citet{Durret2005} for A85, the higher metallicities in those regions cannot be explained by the anti-correlation between metallicity and temperature, they are very likely mergers features. In summary, our findings are consistent with the hypothesis that this cluster is a post-merger, but more detailed studies with simulations are required to verify if the RPS enhancement over the galaxies of group 1 was the responsible for the observed metallicity excess in the western clump. 

\section{SUMMARY AND CONCLUSIONS}

Identifying interactions accurately in optical and X-ray provide us with clues about the merger history of galaxy clusters that help to better understand the influence of extremely violent environments on the evolution of galaxies. A large number of contradictory conclusions about the dynamical state of Abell 2399 was shown in our literature review \citep{Flin2006, Ramella2007,Moretti2017,Bohringer2010,Fogarty2014,Owers2017,Mitsuishi2018}. In this work, we analysed the substructures at different wavelengths and used the results as an input for a hydrodynamical $N$-body simulation. First, we isolated the substructures in the optical data using a Gaussian mixture model code and then we matched them with the X-ray substructures through the comparison with the surface brightness and spectral maps. In the optical analysis, when running \textit{mclust} in 1D, 2D, and 3D, we found that the 3D analysis yielded the most robust results retrieving three separate structures: the main cluster, and two groups. After having isolated those substructures from the optical, we obtained new measures for characteristic radius, mass, and the velocity dispersion of the main cluster and the other two optical groups. In the X-ray analysis, we found the main gas concentration associated with the main cluster and a western clump associated with an optical substructure. Both wavelengths, optical and X-ray, indicated that A2399 is unrelaxed.

The optical group 1, the one that matches the western X-ray clump, has lower pressure, entropy, and temperature than the main cluster, but higher metallicity. We also found that the jellyfish candidate JO69 from \citet{Poggianti2016} is near a shock region of the gas. A possible scenario to explain these characteristics is that this subgroup crossed the main cluster, increasing the ram pressure stripping (that may have influenced on the formation of the jellyfish) in that region and also on the release of metals from the galaxies of that subgroup in the western clump, explaining its higher metallicity.

Using hydrodynamical $N$-body simulations, we offered one specific model that reproduces some of the observed features of A2399 to an approximate degree. The results of the simulation suggest that A2399 could be the outcome of a recent collision taking place on the plane of the sky. This is consistent with the hypothesis put forth by \cite{Mitsuishi2018}, but in the context of our simulation, the temperature discontinuity is understood as a shock front. In our best simulation model, the western clump passed through the main cluster 0.24\,Gyr ago, with a pericentric distance of 110\,kpc. It is currently 580\,kpc ahead of the main cluster core, in an outgoing configuration, and has left behind a $\sim$6\,keV shock front, and also a detached strand of hot gas. The dark matter peak of the western clump has been mostly dissociated from its gas. We propose that this is a plausible scenario to explain the current state of A2399. However, this reconstruction is not unique, even if it is physically well-motivated. Alternative solutions and even an incoming configuration cannot yet be confidently ruled out. Gravitational weak lensing analyses would be needed to map out the dark matter distribution and thus provide more robust constraints for future dedicated simulations.

\section*{ACKNOWLEDGEMENTS}

This research has made use of "Aladin sky atlas" developed at CDS, Strasbourg Observatory, France \citep{2000A&AS..143...33B, 2014ASPC..485..277B}. This research has made use of the VizieR catalogue access tool, CDS, Strasbourg, France. This research made use of XSPEC \citep{Arnaud1996}. This research made use of TOPCAT, an interactive graphical viewer and editor for tabular data \citep{2005ASPC..347...29T}.  This research made use of ds9, a tool for data visualization supported by the Chandra X-ray Science Center (CXC) and the High Energy Astrophysics Science Archive Center (HEASARC) with support from the JWST Mission office at the Space Telescope Science Institute for 3D visualization. Based on observations obtained with XMM-Newton, an ESA science mission with instruments and contributions directly funded by ESA Member States and NASA. This work made use of the computing facilities of the Laboratory of Astroinformatics (IAG/USP, NAT/Unicsul), whose purchase was made possible by the Brazilian agency FAPESP (grant 2009/54006-4) and the INCT-A. Also, this work made use of the computing facilities at the Centro de Computa\c c\~ao Cient\'ifica e Tecnol\'ogica (UTFPR). 
ACCL thanks the financial support of the National Agency for Research and Development (ANID) / Scholarship Program / DOCTORADO BECAS CHILE/2019-21190049. ACCL also acknowledges the financial suport of the Coordenaa\c c\~ao de Aperfei\c coamento de Pessoal de N\'ivel Superior -- Brasil (CAPES) -- Finance Code 001.
PAAL thanks the support of CNPq, grant 309398/2018-5. 
TFL acknowledges financial support from FAPESP (2018/02626-8) and CNPq (303278/2015-3). 
REGM acknowledges support from the Brazilian agency \textit {Conselho Nacional de Desenvolvimento Cient\'ifico e Tecnol\'ogico} (CNPq) through grants 303426/2018-7 and 406908/2018-4.
Y.J. acknowledges financial support from CONICYT PAI (Concurso Nacional de Inserci\'on en la Academia 2017) No. 79170132 and FONDECYT Iniciaci\'on 2018 No. 11180558.
ALBR thanks for the support of CNPq, grant 311932/2017-7.
BV and AM acknowledge the financial contribution from the contract ASI-INAF n.2017-14-H.0, from the grant PRIN MIUR 2017 n.20173ML3WW\_001 (PI Cimatti) and from the INAF main-stream funding programme (PI Vulcani).
RSN thanks the financial support from CNPq, grant 301132/2020-8. The authors acknowledge the National Laboratory for Scientific Computing (LNCC/MCTI, Brazil) for providing HPC resources of the SDumont supercomputer, which have contributed to the research results reported within this paper.

\section*{Data availability}
The data underlying this article were derived from sources in the public domain and are available in Vizier, at \url{https://dx.doi.org/10.1051/0004-6361:200810997} and \url{https://dx.doi.org/10.1051/0004-6361/201630030}. 






\bibliographystyle{mnras} 
\bibliography{refs2}



\appendix

\section{Goodness of fitting}
\label{appendix:a}

A probability distribution that consists of multiple Gaussians depends on a co-variance matrix $\Sigma_{k}$ which can be understood as the variance in multiple dimensions. The eigen-decomposition of the form $\bm{\Sigma}_{k} = \lambda_{k} \bm{D}_{k} \bm{A}_{k} \bm{D}_{k}^\top$ provides parsimonious parametrisations of the  co-variance matrices for each model fitted by \textit{mclust}. The volume of the ellipsoid described by $\bm{\Sigma}_{k}$ is regulated by the scalar $\lambda_{k}$, the shape of the isodensity contours is outlined by the diagonal matrix $\bm{A}_{k}$, and the ellipsoid orientation is determined by the orthogonal matrix $\bm{D}_{k}$ of the $k\rm^{th}$ component.

The parameters of the co-variance matrices defining each different model fitted by \textit{mclust} are described in the following table. Each letter in the model identifier column encodes a different meaning according to its position. The first letter gives information about the volume, the second describes the shape, and the third specifies the orientation of the multidimensional Gaussians. The letters used for describing those features are E (equal), V (varying across clusters), and I (related to the identity matrix in specifying shape or orientation). In a model with identifier VEI, for example, the retrieved clusters will present varying volumes, equal shapes, and the identity I as the orientation. \citep{fraley2012}. 

\begin{table*}
\caption{Models adjusted by \textit{mclust}. The `x' in the third and fourth columns represents the availability in the column Hierarchical Cluster (HC) and Expectation-Maximisation (EM). Adapted from \citet{Scrucca_2016}.}
\begin{center}
\begin{tabular}{ |p{1.0cm}|p{1.3cm}|p{1.0cm}|p{1.0cm}|p{2,0cm}|p{1.3cm}|p{1.3cm}|p{2.2cm}| }
 \hline\hline
 \multicolumn{8}{|c|}{Models adjusted by \textit{mclust}} \\
 Id & $\Sigma_{k}$ & HC & EM & Distribution & Volume & Shape & Orientation \\
 \hline
 E  & & x& x& (univariate)& Equal& & \\
 
 V  & & x& x& (univariate)& Variable& & \\
 
 EII & $\lambda\textbf{\textit{I}}$ & x& x& Spherical& Equal& Equal&NA\\
 
 VII &  $\lambda_{k}\textbf{\textit{I}}$ & x&x& Spherical& Variable& Equal&NA\\
 
 EEI & $\lambda\textbf{\textit{A}}$ & &x& Diagonal& Equal& Equal& Coordinate axes\\
 
 VEI & $\lambda_{k}\textbf{\textit{A}}$ & & x& Diagonal& Variable& Equal&  Coordinate axes\\
 
 EVI &  $\lambda\textbf{\textit{A}}_{k}$ & & x& Diagonal& Equal& Variable& Coordinate axes\\
 
 VVI & $\lambda_{k}\textbf{\textit{A}}_{k}$ & & x& Diagonal& Variable& Variable& Coordinate axes\\
 
 EEE & $\lambda\textbf{\textit{D}}\textbf{\textit{A}}\textbf{\textit{D}}^\top$ & x& x& Ellipsoidal& Equal& Equal& Equal\\
 
 EVE &  $\lambda\textbf{\textit{D}}\textbf{\textit{A}}_{k}\textbf{\textit{D}}^\top$ & & x& Ellipsoidal& Equal& Variable& Equal\\
 
 VEE & $\lambda_{k}\textbf{\textit{D}}\textbf{\textit{A}}\textbf{\textit{D}}^\top$ & & x& Ellipsoidal& Variable& Equal& Equal\\
 
 VVE & $\lambda_{k}\textbf{\textit{D}}\textbf{\textit{A}}_{k}\textbf{\textit{D}}^\top$ & & x& Ellipsoidal& Variable& Variable& Equal\\ 
 
 EEV & $\lambda\textbf{\textit{D}}_{k}\textbf{\textit{A}}\textbf{\textit{D}}_{k}^\top$ & & x& Ellipsoidal& Equal& Equal& Variable\\
 
 VEV & $\lambda_{k}\textbf{\textit{D}}_{k}\textbf{\textit{A}}\textbf{\textit{D}}_{k}^\top$ & & x& Ellipsoidal& Variable& Equal& Variable\\
 
 EVV & $\lambda\textbf{\textit{D}}_{k}\textbf{\textit{A}}_{k}\textbf{\textit{D}}_{k}^\top$ & & x& Ellipsoidal& Equal& Variable& Variable\\
 
 VVV & $\lambda_{k}\textbf{\textit{D}}_{k}\textbf{\textit{A}}_{k}\textbf{\textit{D}}_{k}^\top$ & x& x& Ellipsoidal& Variable& Variable& Variable\\
 \hline

\end{tabular}

\end{center}
\label{tbl:tabular}
\end{table*}

\bsp	
\label{lastpage}
\end{document}